\documentclass[useAMS,usenatbib,usegraphicx]{mn2e}
\usepackage{epsf,url}
\usepackage{subfigure,amssymb,amsfonts,amstext,amsgen,amsopn,amsxtra,indentfirst,lscape,times,rotating}
\usepackage{longtable}
\usepackage{pdflscape}
\usepackage{graphics}
\usepackage{keyval}
\usepackage{trig}
\usepackage[dvips]{color}
\usepackage{dcolumn}
\usepackage{bm}
\def\beq{\begin{equation}}
\def\eeq{\end{equation}}
\def\bey{\begin{eqnarray}}
\def\eey{\end{eqnarray}}
\def\msun{M_\odot}

\def\kms{\, {\rm km \, s}^{-1} }

\def\prd{Phys. Rev. D}
\def\mnras{MNRAS}
\def\apj{ApJ}

\def\apjl{ApJ}

\def\aap{A \& A}
\def\aj{AJ}

\def\aap{Astron. Astrophys.}

\title{A QUMOND galactic N-body code I: Poisson solver and rotation curve fitting}

\author[G. W. Angus et al.]
{\parbox{\textwidth}{G.~W.~Angus$^{1}$\thanks{E-mail: \texttt{angus.gz@gmail.com}},
K. van der Heyden$^{1}$,
B. Famaey$^{2,3}$,
G. Gentile$^{4}$,
S. S. McGaugh$^{5}$ and 
W. J. G. de Blok$^{1}$}\vspace{0.4cm}\\
\parbox{\textwidth}{
$^{1}$Astrophysics, Cosmology \& Gravity Centre, University of Cape Town, Private Bag X3, Rondebosch, 7701, South Africa \\
$^{2}$Observatoire Astronomique de Strasbourg, CNRS, UMR 7550, France  \\
$^{3}$AIfA, University of Bonn, Germany \\
$^{4}$Sterrenkundig Observatorium, Ghent University, Krijgslaan 281, S9, 9000 Ghent, Belgium \\
$^{5}$Department of Astronomy, Computer \& Space Sciences Bldg, University of Maryland, College Park, MD 20742-2421, USA}}
\begin{document}

\date{\today}
\maketitle
\begin{abstract}
Here we present a new particle-mesh galactic N-body code that uses the full multigrid algorithm for solving the modified Poisson equation of the Quasi Linear formulation of Modified Newtonian Dynamics (QUMOND). A novel approach for handling the boundary conditions using a refinement strategy is implemented and the accuracy of the code is compared with analytical solutions of Kuzmin disks. We then employ the code to compute the predicted rotation curves for a sample of five spiral galaxies from the THINGS sample.  We generated static N-body realisations of the galaxies according to their stellar and gaseous surface densities and allowed their distances, mass-to-light ratios ($M/L$) and both the stellar and gas scale-heights to vary in order to estimate the best fit parameters. We found that NGC~3621, NGC~3521 and DDO~154 are well fit by MOND using expected values of the distance and $M/L$. NGC~2403 required a moderately larger $M/L$ than expected and NGC~2903 required a substantially larger value. The surprising result was that the scale-height of the dominant baryonic component was well constrained by the rotation curves: the gas scale-height for DDO~154 and the stellar scale-height for the others. In fact, if the suggested stellar scale-height (one-fifth the stellar scale-length) was used in the case of NGC~3621 and NGC~3521 it would not be possible to produce a good fit to the inner rotation curve. For each of the four stellar dominated galaxies, we calculated the vertical velocity dispersions which we found to be, on the whole, quite typical compared with observed stellar vertical velocity dispersions of face on spirals. We conclude that modelling the gas scale-heights of the gas rich dwarf spiral galaxies will be vital in order to make precise conclusions about MOND. 
\end{abstract}

\section{Introduction}
\protect\label{sec:intr}
The {H\,{\sc I~}} Nearby Galaxy Survey (THINGS, \citealt{walter08}) brought an unprecedented level of precision to the measurement of the rotation curves of certain nearby spiral galaxies. This, when coupled with 3.6~$\mu m$ images of the stellar component from the Spitzer Infrared Nearby Galaxies Survey (SINGS, \citealt{kenn03}), produces a stringent new data set for modeling the galactic dynamics of these systems. These tighter constraints are of paramount importance for testing alternative theories of gravity, in particular those with no galactic dark matter like Modified Newtonian Dynamics (MOND - \citealt{milgrom83a}, but see also \citealt{famaey12}).

MOND is an appealing framework because, at least in galaxies, the gravitational field is fully determined by the matter distribution of the visible components. This means that a galaxy comprised of disky stellar and gaseous components, produces a rotation curve depending only on the properties of those components. It is crucial in order to keep up with the advancing observations, that we produce methods of modelling the galaxies with a similar level of sophistication.

In this article, we develop a galactic N-body code for MOND and apply it to fitting the high resolution rotation curves of the THINGS survey. We compare predicted and fitted distances and make use of free parameters in the form of the scale-heights of the two baryonic components.

\section{Solving the MOND equation on a grid}

MOND has a different force law from Newtonian dynamics, and this traditionally is rather tricky to solve (\citealt{bekenstein84,milgrom86,brada95}). For instance, in the Newtonian analogue, the ordinary Poisson equation must simply be solved using the matter distribution in terms of its density,  which includes stars ($\rho_{*}$), gas ($\rho_g$) and cold dark matter ($\rho_{CDM}$), using

\beq
\protect\label{eqn:qumond1}
\nabla^2\Phi_N=4\pi G (\rho_{*}+\rho_g+\rho_{CDM})
\eeq
to give the Newtonian potential, $\Phi_N$. However, in a recently proposed version of MOND, called QUMOND (\citealt{milgrom10}, but see also \citealt{zhao10}) it was shown that the MOND potential, $\Phi$, can be found exactly from the Newtonian potential (not including cold dark matter) as follows
\beq
\protect\label{eqn:qumond2}
\nabla^2\Phi=\nabla \cdot \left[ \nu(y) \nabla\Phi_N \right],
\eeq
where $\nu(y)=0.5+0.5\sqrt{1+4/y}$ and $y=\nabla\Phi_N/a_o$, with $a_o$ being the MOND acceleration constant chosen here to be $3.6~(\kms)^2pc^{-1}$. $\nu(y)$ could take on another form, and similarly $a_o$ could take on a different value, but we choose here not to focus on this topic for fear of being sidetracked. We note that the $\nu$-function we adopt here is the analogue of the $\mu$-function used by \cite{fb05} to fit the terminal velocity curve of the Milky Way.

Very often in the literature, Eq~\ref{eqn:qumond2} is not solved, but rather an approximation is made that is referred to as the algebraic MOND relation

\beq
\protect\label{eqn:algmond}
\nabla\Phi=\nu(y) \nabla\Phi_N.
\eeq
This approximation ignores the curl field, which is negligible in most cases of high symmetry, but has the scope to be a significant component of the gravitational field in triaxial systems. Here we are not advocating that it is imperative to use Eq~\ref{eqn:qumond2} for fitting rotation curves of relaxed spiral galaxies, but rather that when MOND is extended to study interacting galaxies, or galaxies with lopsided matter distributions.

Our goal is to determine the gravity at locations inside the galaxy we are studying and the procedure to do this is as follows. We must set up a three dimensional grid, sliced into numerous cubic cells - usually we use 65, 129 or 257 cells per dimension. In each cell there are the following.

\begin{itemize}
 \item An approximated, but near exact, value of the baryonic density (step 1, the source of Eq~\ref{eqn:qumond1} using only gas and stars).
 \item A solution to the Newtonian potential (step 2).
 \item The source ``density'' - the right hand side of Eq~\ref{eqn:qumond2} (step 3).
 \item The solution of the QUMOND potential (step 4)
\end{itemize}

Getting from step 1 to 4 is in the details of the code, which we briefly describe here.

We read in a set of particle positions that determine the density of the stars and gas in the galaxy on the grid using the cloud-in-cell technique (step 1). The stars and gas are allocated half of the total number of particles each and therefore the particles of gas and stars are weighted differently and according to the total mass of each component. With this density, we solve for the Newtonian potential (step 2; Eq~\ref{eqn:qumond1}) using the full multigrid algorithm that is well known and described in \cite{numrec}. The full multigrid method is extended to three dimensions which implements red-black Gauss-Seidel relaxation. Our default is to use six V-cycles, from finest to coarsest grid and back to finest with bilinear prolongation and restriction, as well as two pre and post smoothing steps before and after the coarse grid correction is computed.

We then use finite differencing to find the source for Eq~\ref{eqn:qumond2} (step 3). The finite differencing technique used to solve Eq~\ref{eqn:qumond2} is described in \cite{angus11}. We use the same multigrid technique to solve Eq~\ref{eqn:qumond1} and Eq~\ref{eqn:qumond2}, only the source changes (step 4). We make a final finite differencing to determine the QUMOND gravity, $\nabla\Phi$, on the grid and then we interpolate to the point at which we wish to know the gravity. The interpolation scheme is the same as the scheme to assign density to the grid i.e. the cloud-in-cell technique.

Our final step is to calculate the circular speed associated with this QUMOND potential as a function of radius. This is taken as 
\beq
\protect\label{eqn:vc}
v_c^2=R\times\partial_R\Phi|_{z=0}.
\eeq
 Although this is the first attempt to solve the modified Poisson equation of QUMOND on a grid in a non-cosmological setting (see \citealt{llinares08,llinares11,angus11} for the cosmological analogue), this is not the first time a modified Poisson equation linked to MOND has been solved in such a way. \cite{brada99b} developed a cubic particle-mesh code that solved the modified Poisson equation and used it to investigate some important topics (\citealt{brada00a,brada00b}). \cite{tcevol} developed a similar code and parallelised it and incorporated hydrodynamics into their simulations (see \citealt{tc08,tiret08a}). In addition to these, \cite{londrillo09} produced a spherical grid code that investigated various issues in the framework of MOND (\citealt{nipoti07a,nipoti07c,nipoti08}).

The key to a galactic N-body code, as opposed to a cosmological one, is the boundary conditions. In a cosmological code, the expectation that the Universe is homogeneous and isotropic allows us to enforce periodic boundary conditions. At the galaxy scale, the boundary conditions must be set precisely at all outer grid cells, which is non-trivial in MOND. It is not clear how accurately this was achieved in the previous codes, but here we implement a different strategy (see \citealt{wu09} and \citealt{tcevol} for other examples).

We define a coarse grid which is many times ($2^{10}$) larger than the galaxy we are studying. For this coarsest grid we set the boundary condition of the Newtonian and QUMOND potentials to be zero. Then we solve for the Newtonian and QUMOND potentials everywhere on the grid. From this, we define a refined grid that is half the size of the coarse grid and we interpolate through the values of the potential on the coarse grid to define the boundary condition of the refined grid. We make this refinement up to ten times in order to zoom in on our galaxy, but the boundary conditions are correct at the sub percent level at all points on the grid by the second refinement. This also enables us to use different grid levels to find the potential at the location of a particle, depending on its position in the various grid levels. The limitation of this is that it is centrally refining, not arbitrarily towards regions of high density. Note that the number of grid cells is constant for each refinement, so the resolution increases for the finer grids - with more cells per kpc.

There is an initial guess that is required in multigrid methods. For clarity, our finest grid is 257x257x257 and we solve for the potential on this grid at 10 incremental levels of resolution (box size). For each box size, in order to more rapidly calculate the potential on the finest grid, we smooth the calculated (fed in) density down to the coarsest possible level of resolution, which is a 3x3x3 grid. The coarsest density field is known in each of the 27 cells and we must solve for the potential in each of these cells. Since for the largest box, of length 4~Mpc, the density and potential in each of the 26 outer cells is zero, the potential in the central cell is simply proportional to the density in that central cell. This is true for both the Newtonian step and the QUMOND step. Following this, we perform the standard multigrid interpolation of the potential to finer grids and use the finer density defined on those grids to make a more rapid calculation of the potential on the finest grid.

\section{Static N-body realisations}
\protect\label{sec:ics}
Our aim is to use this particle mesh code to fit rotation curves. To do this, we need to generate static N-body realisations of galaxies depending on their surface density. Initially, to test the resolution of the code, we generated a realisation of a Kuzmin disk and compared the analytical circular velocity to the circular velocity simulated by the code.

\subsection{Kuzmin Disks}
Kuzmin disks are unique in MOND because their isopotential contours are spherically symmetric (at least in each distinct hemisphere, above and below the disk mid-plane), meaning there is no curl field. The surface density of a Kuzmin disk is given by \cite{bt08} to be $\Sigma_K(R)={aM \over 2\pi}(R^2+a^2)^{-3/2}$. Assuming an infinitely thin disk, and integrating this by $2\pi RdR$, we get the enclosed mass $M_K(R)=M(1-{a \over \sqrt{R^2+a^2}})$. If we express the fraction of the total mass as $f=M_K(R)/M$, then we know the way to distribute particles in radius is $R=a\sqrt{(1-f)^{-2}-1}$. From this we randomly sample $f$ and distribute the particles according to a random azimuthal angle.

\begin{figure*}
\centering
\subfigure{
\includegraphics[angle=0,width=8.50cm]{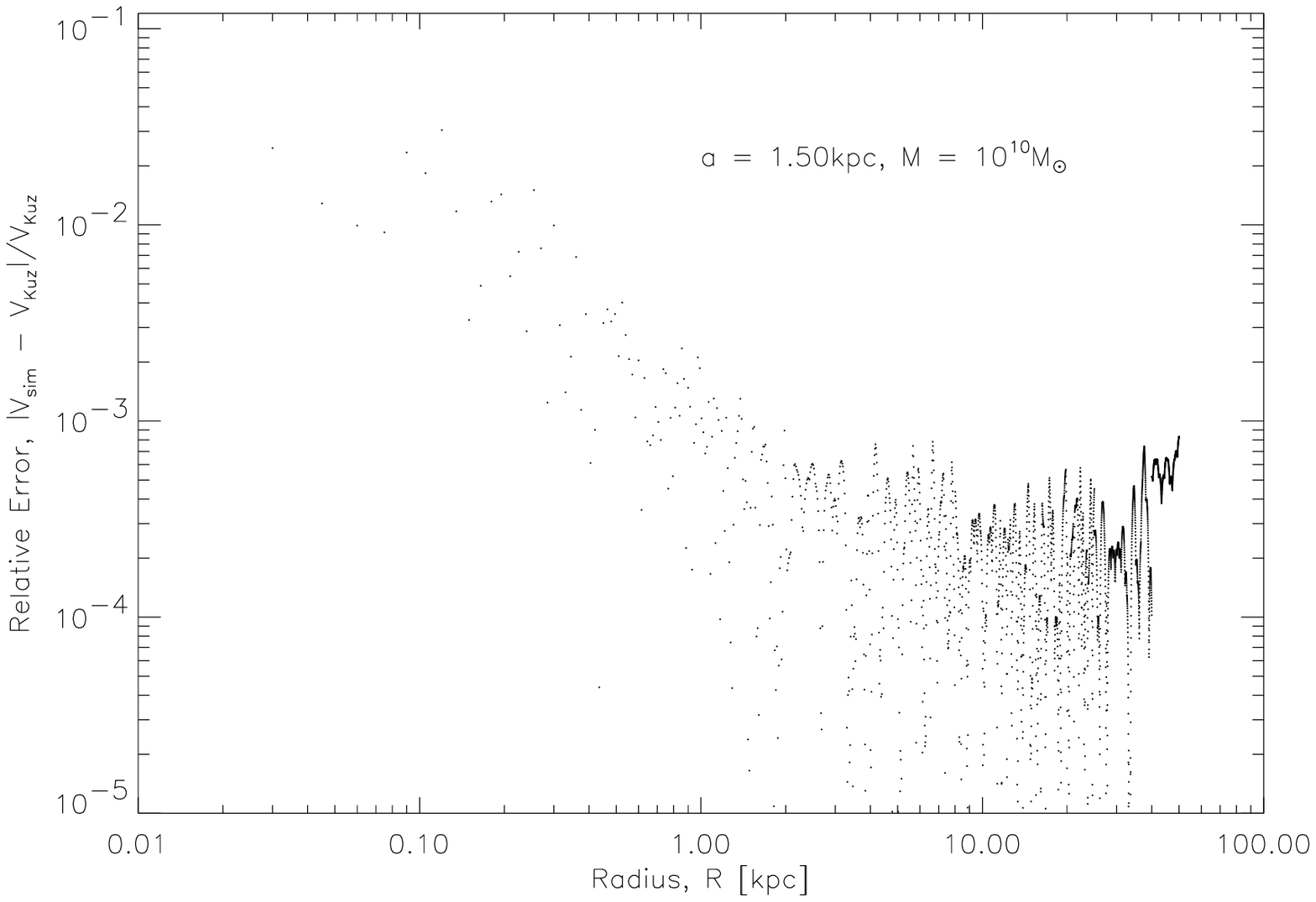}
}
\subfigure{
\includegraphics[angle=0,width=8.50cm]{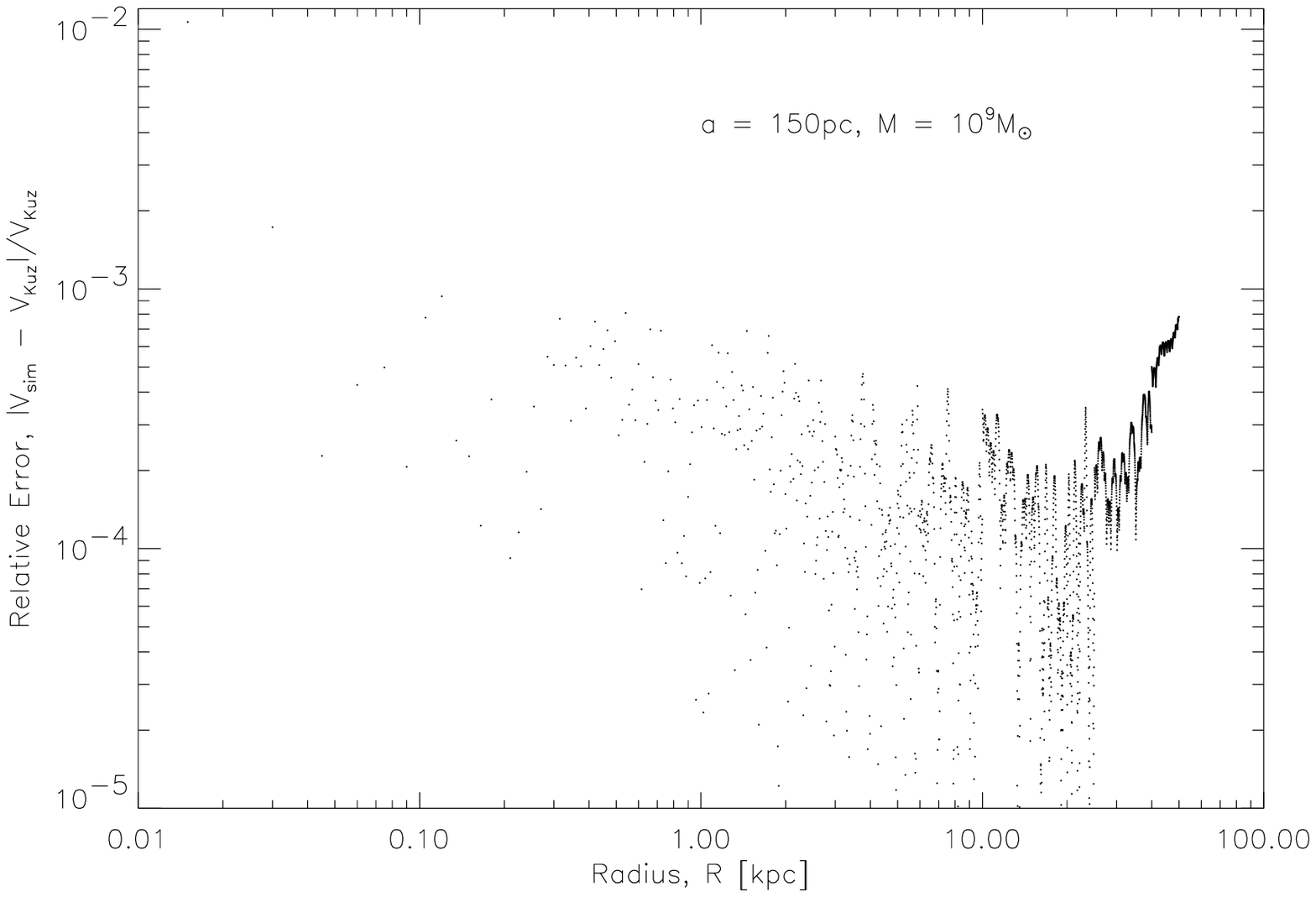}
}\\
\caption{Shows the relative error between the analytically defined QUMOND circular velocity and the simulated circular velocity of a Kuzmin disk. The left hand panel uses a mass of $10^{10}\msun$ and scale-length of $1.5$~kpc, whereas the right hand panel uses a mass of $10^{9}\msun$ and scale-length of $150$~pc. Note that the y-axes have a different ranges.}
\label{fig:kuz}
\end{figure*}

We did this for a model where we set $a=1.5$~kpc and $M=10^{10}\msun$ and we show in Fig~\ref{fig:kuz} (left hand panel) the fractional difference between this, the simulated QUMOND rotation curve, and the analytically expected rotation curve. For the Kuzmin disk, the 3D Newtonian gravity on a star slightly below the disk is defined by $g_{N,r}=-GM/(R^2+a^2)$ and this points towards the position $(R,z)=(0,a)$, as if all the mass was concentrated in a single point at the location $(0,a)$. Similarly, the gravity towards the centre of the disk is $g_{N,R}=-GMR(R^2+a^2)^{-3/2}$. Therefore, the QUMOND gravity towards the centre of the disk is $g_{M,R}=\nu(|g_{N,r}|/a_o)g_{N,R}$. Note the subtlety here with respect to the subscripts of R and r.

Fig~\ref{fig:kuz} demonstrates that the accuracy of the code is less than 1\% beyond 0.3~kpc and less than 0.1\% beyond 1~kpc for this model. This depends greatly on the extent of the galaxy. A Kuzmin disk, with scale radius $a=1.5$~kpc still has 5\% of its mass beyond 30~kpc and so is rather unrealistically extended. In Fig~\ref{fig:kuz} (right hand panel) we make the same plots for $a=150$~pc and $M=10^{9}\msun$ to demonstrate that in that case the agreement is better than 0.1\% down to $\sim$50~pc. In both cases, as for the real galaxies to follow, we use the following strategy to evaluate gravity at various radii, R. When $R < 0.8~kpc$ we use the finest grid which has a default box-size of 4~kpc, when $0.8<R<1.6~kpc$ we use the second finest box of size 8~kpc, for $1.6<R<3.2~kpc$ we use the 16~kpc box and so on. These values need not be fixed and should be tailored to different galaxy sizes.

\subsection{Convergence}
Although we can accurately simulate the gravitational field of a Kuzmin disk to roughly 0.1\%, it is equally important to ensure that the solution to Poisson's equation does converge for more complex density distributions, which might not immediately follow from a smooth Kuzmin disk. As mentioned above, our default parameters for using the full multigrid algorithm are that we make a standard number of six V-cycles at each level of the grid. Before and after we make the coarse grid correction we always make two red-black Gauss-Seidel relaxation sweeps.

\begin{figure}
\centering
\subfigure{
\includegraphics[angle=0,width=8.50cm]{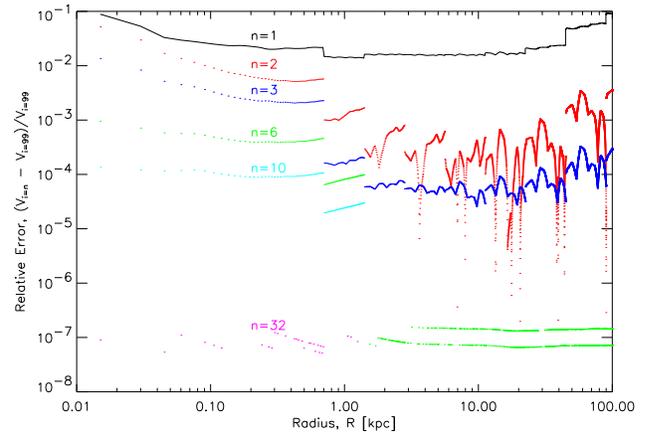}
}
\caption{For NGC 3621, shows the relative error (as a function of radius) between the simulated rotation speed for a series of numbers of V-cycles compared to the rotation speed found using 99 V-cycles. Each different number of V-cycles is given a distinct colour. If a colour is not seen at a specific radius, this is because the relative error is less than $10^{-8}$. The discontinuities are caused by moving from one grid resolution to another.}
\label{fig:converge}
\end{figure}

For one of our galaxies, NGC 3621, we have plotted in Fig \ref{fig:converge} the relative error (as a function of radius) between the simulated rotation speed for a series of numbers of V-cycles compared to the rotation speed found using 99 V-cycles. Each different number of V-cycles is given a distinct colour. If a colour is not seen at a specific radius, this is because the relative error is less than $10^{-8}$ and is most likely identical to the 99 V-cycles simulation. The discontinuities are caused by moving from one grid resolution to another. Clearly, using only one or two V-cycles does not allow for convergence, but by six cycles the difference is less than 0.1\% at all radii. Since the theoretical accuracy of the code is roughly 0.1\% as well (see Fig \ref{fig:kuz}), it makes 6 V-cycles a sensible number of cycles to make until the accuracy can be improved.

\subsection{Curl-field and scale-height}
\begin{figure}
\centering
\subfigure{
\includegraphics[angle=0,width=8.50cm]{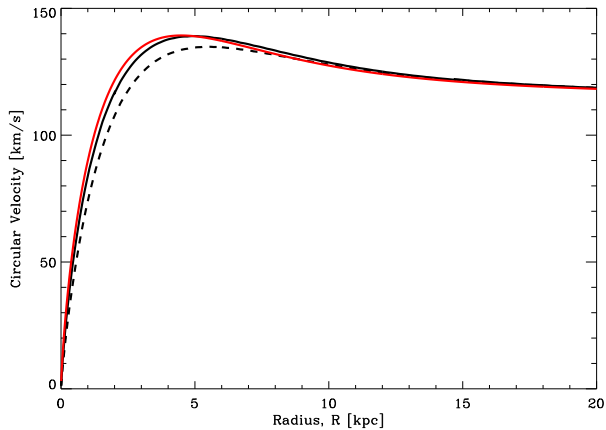}
}
\caption{Shows the difference between the rotation curve when we solve the modified Poisson equation of QUMOND (solid black, Eq~\ref{eqn:qumond2}) and the algebraic relation (solid red, Eq~\ref{eqn:algmond}) for a razor thin exponential disk. The dashed curve shows the rotation speed from the same exponential disk, using our code, but with a vertical scale-height of 1~kpc that is described by a $sech^2$ distribution.}
\label{fig:exp}
\end{figure}

To compare the difference between solving the modified Poisson equation of QUMOND (Eq~\ref{eqn:qumond2}) and the algebraic MOND relation (Eq~\ref{eqn:algmond}) we plot in Fig~\ref{fig:exp} the rotation curve for an exponential disk galaxy with scale length $a=1.5$~kpc and total mass $M=10^{10}\msun$. There is a small but noticeable difference even for a perfectly symmetric and thin exponential disk. Also plotted is the simulated rotation curve for a disk with scale-height of 1~kpc, with a $sech^2$ distribution. This creates a $10~\kms$ difference out to $R=6$~kpc and is still significant at $R=10$~kpc. For this reason, scale-height is a crucial parameter that we will use to enhance the fits to the rotation curves of the galaxies.

\subsection{Realistic two-component galaxies}
\begin{figure}
\centering
\subfigure{
\includegraphics[angle=0,width=8.50cm]{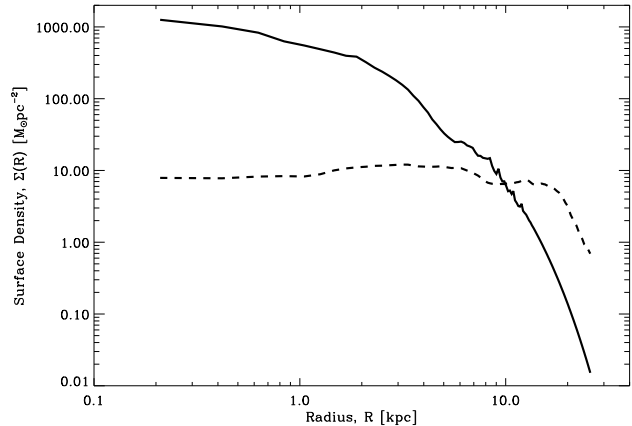}
}
\caption{Shows the surface densities of the stellar (solid) and gaseous (dashed) components for NGC 3621. From de Blok et al. (2008).}
\label{fig:surfd}
\end{figure}

In order to generate realistic, static N-body realisations for galaxies we need the surface densities of both the stellar disk and the gaseous disk, like that plotted for NGC 3621 in Fig~\ref{fig:surfd}. In our models we always separate stars and gas, giving half of the particles to each component and weighting them according to the relative masses of these two components. We use the well known rejection technique from \cite{numrec} to produce an N-body realisation of the two components that resembles the observed surface densities. Typically we use $256^3$ particles and our most refined grid is only 4~kpc across, with $129$ cells per dimension.

\subsection{Free parameters}
We are now in a position to compare simulated circular velocity curves with the observed rotation curves. This brings into question the various free parameters we employ. We assume that the uncertainty in the inclination of the galaxy is contained within the measurement errors associated with each data point. The error from inclination is addressed in de Blok et al. (2008), where the tilted ring fits find little variation in inclination for the galaxies we use. This leaves us with four free parameters: the distance to the galaxy, $D$, the mass-to-light ratio ($M/L$) of the stellar component of the galaxy and the scale-heights of both the stellar, $z_*$, and gas, $z_g$, disks - with $sech^2$ distributions. We then make an exhaustive search through parameter space to find the quality of fit values, $\chi^2/n=\left( \Sigma_{i=1}^n {(V_{fit}(R_i) - V_{obs}(R_i))^2 \over \sigma^2_{obs}(R_i)}\right)/n$, of these parameters by comparing the simulated QUMOND rotation curve with the observed one. We initially enforce flat priors on the distance to be no more than two standard deviations and the $M/L$ is forced to be within the range set by the diet-Salpeter and Kroupa IMFs. We then relax these priors if no good fit is found, or if a significantly improved fit is found despite not being preferred by current limits on the parameters. The scale-heights are free to vary from razor thin up to 1~kpc. It should be emphasised that \cite{gentile11} have suggested that \cite{deblok08} overestimate the error bars on some rotation curves, $\sigma_{obs}$, and as such the reduced $\chi^2$ cannot be directly used as a probability indicator in every case.

\section{Sample}
The sample of five galaxies that we've chosen is a subsample of the galaxies used in the MOND fits of \cite{gentile11}, which is itself a subset of the galaxies studied by \cite{deblok08}. Our sample removed all galaxies that had a significant spheroid component and also those galaxies that could be aided by the external field effect, which we devote a future paper to. Here our goal is not to have a representative sample of galaxies to test MOND in a statistical sense, but rather to highlight the potential of this Poisson solver and to investigate the impact of variable scale-heights on rotation curve fits. The properties of the five galaxies themselves can be found detailed in \cite{deblok08}.

\section{Results}
\begin{figure*}
\centering
\subfigure{\label{fig:dist}
\includegraphics[angle=0,width=8.50cm]{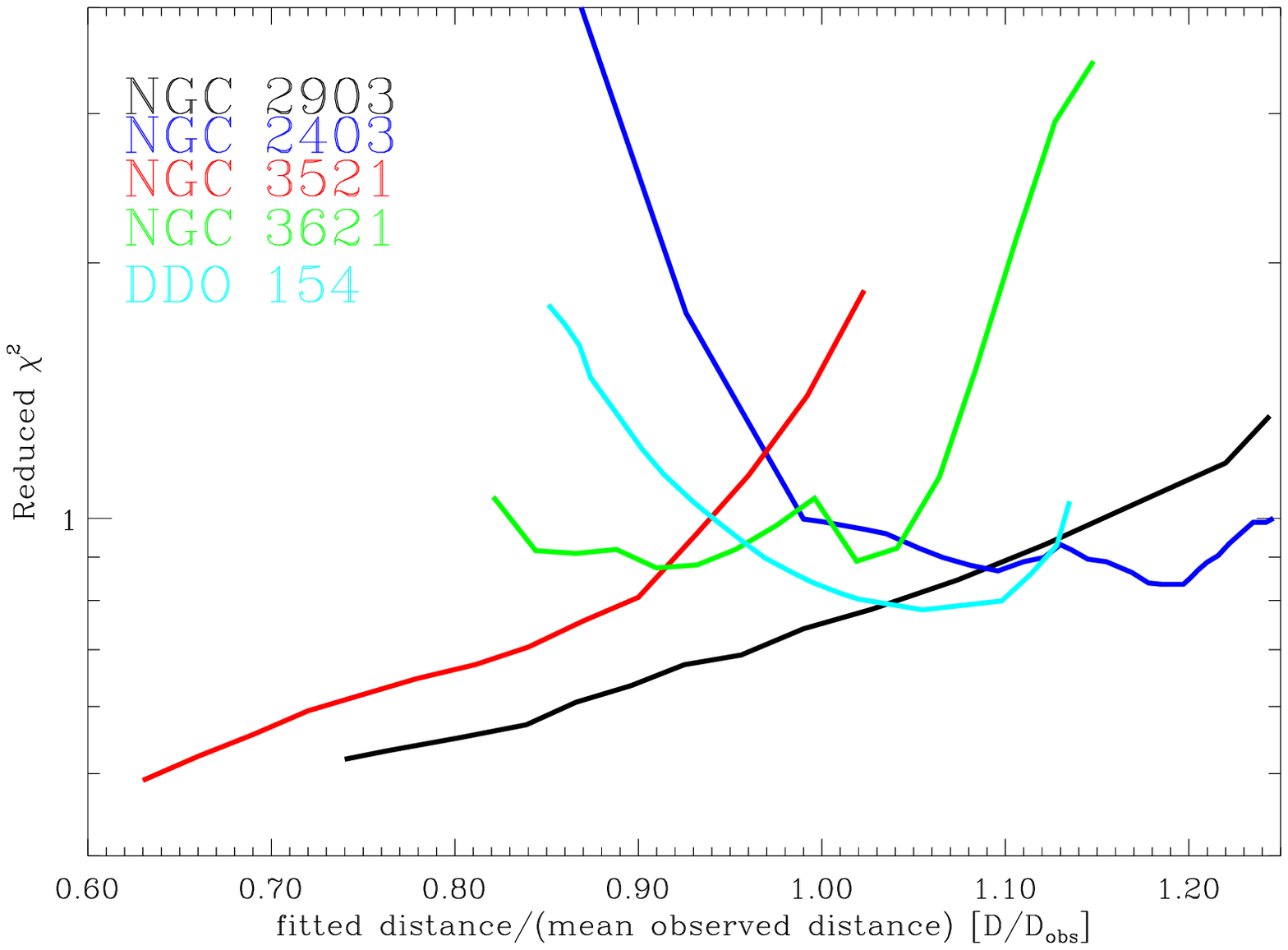}
}
\subfigure{\label{fig:m2l}
\includegraphics[angle=0,width=8.50cm]{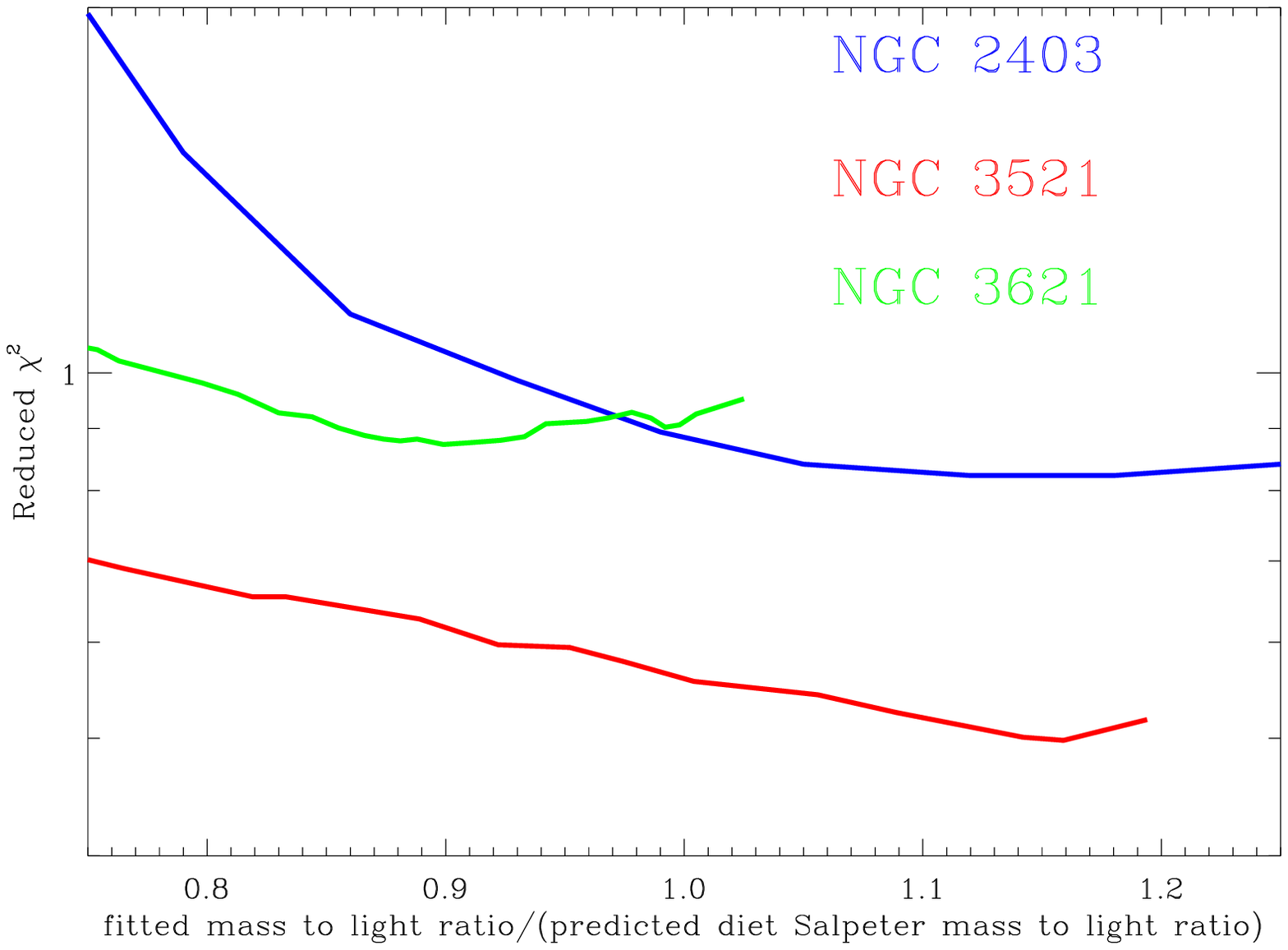}
}\\
\subfigure{\label{fig:dist1}
\includegraphics[angle=0,width=8.50cm]{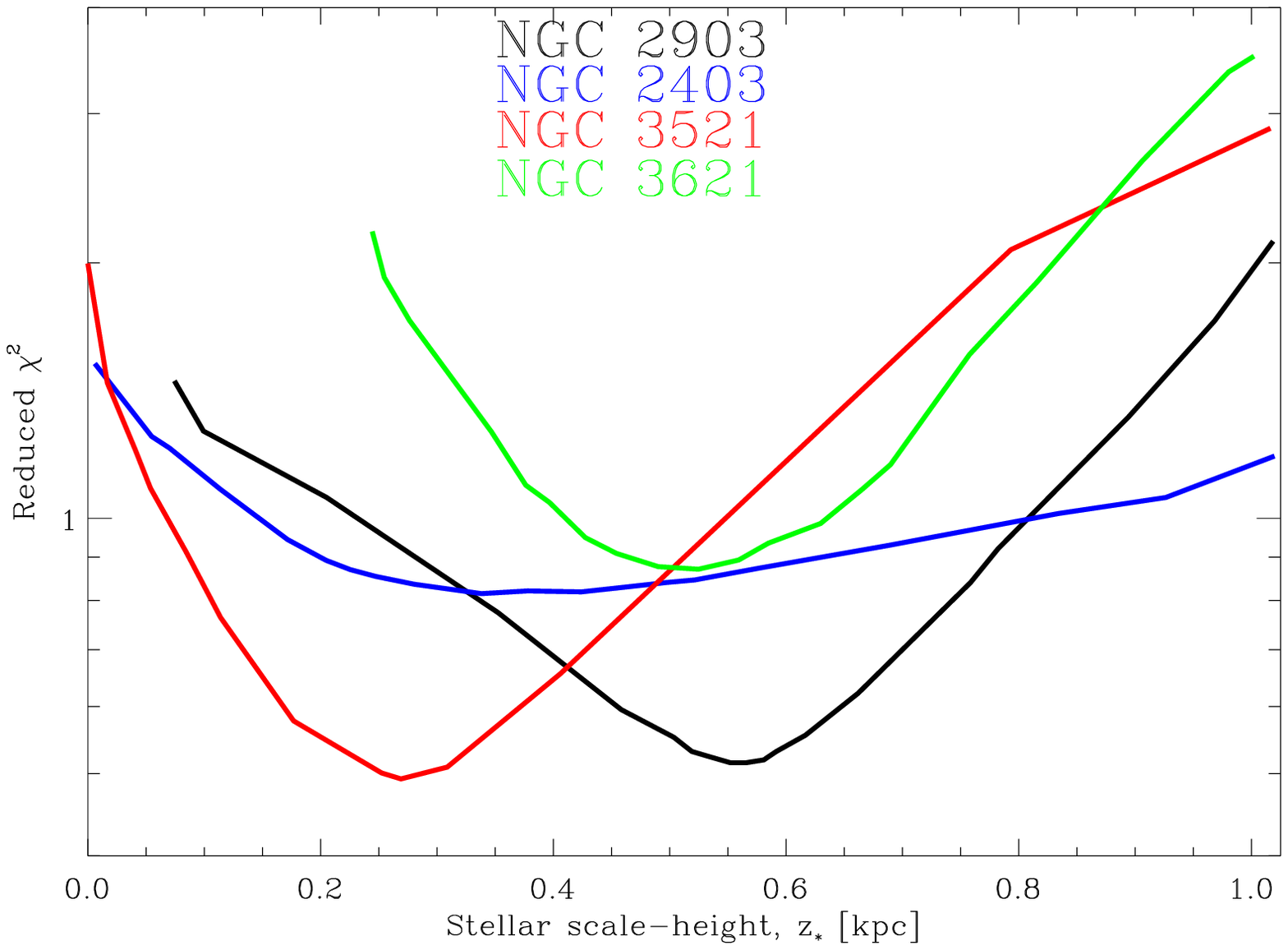}
}
\subfigure{\label{fig:m2l1}
\includegraphics[angle=0,width=8.50cm]{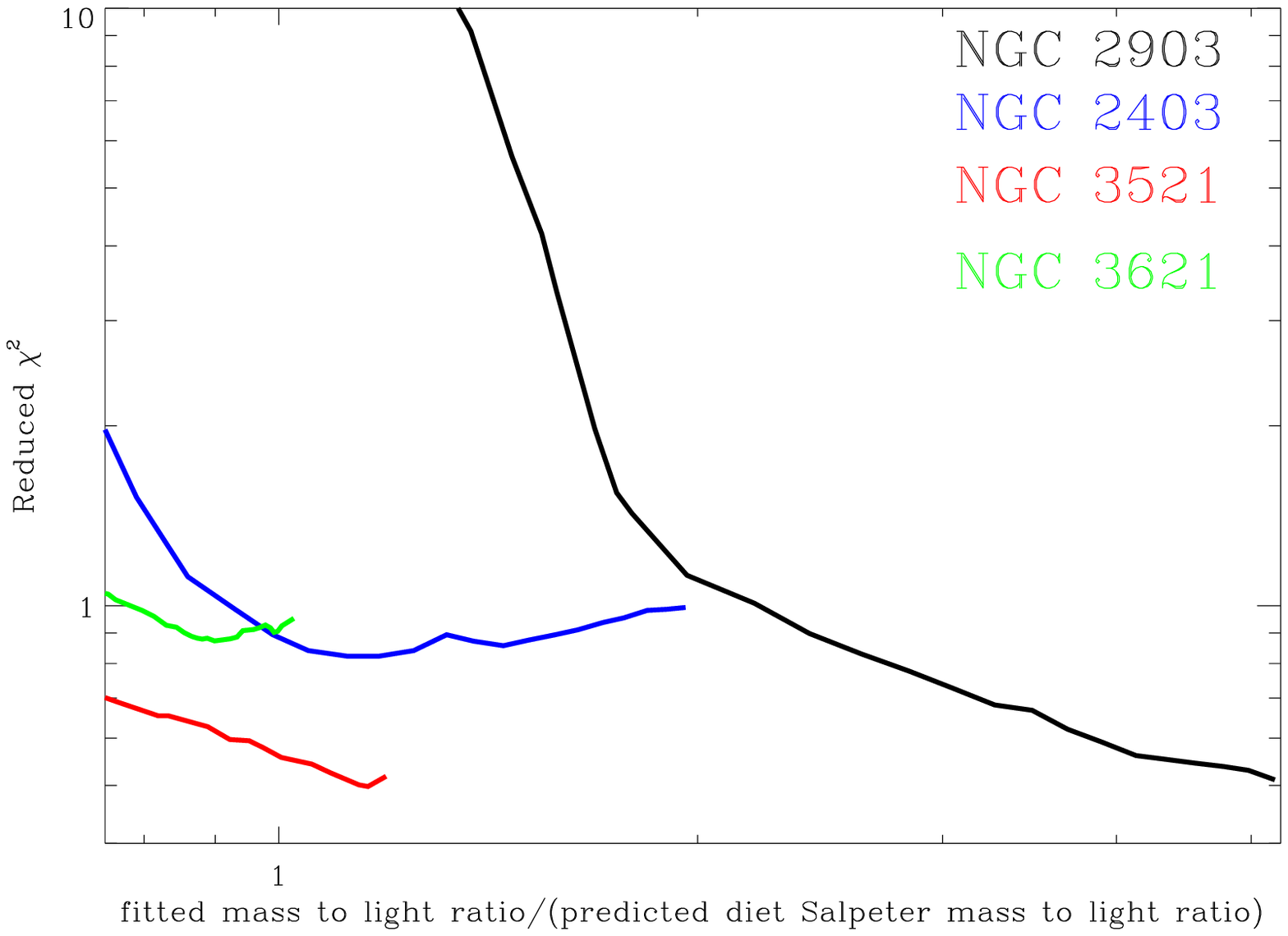}
}\\
\subfigure{\label{fig:m2l2}
\includegraphics[angle=0,width=8.50cm]{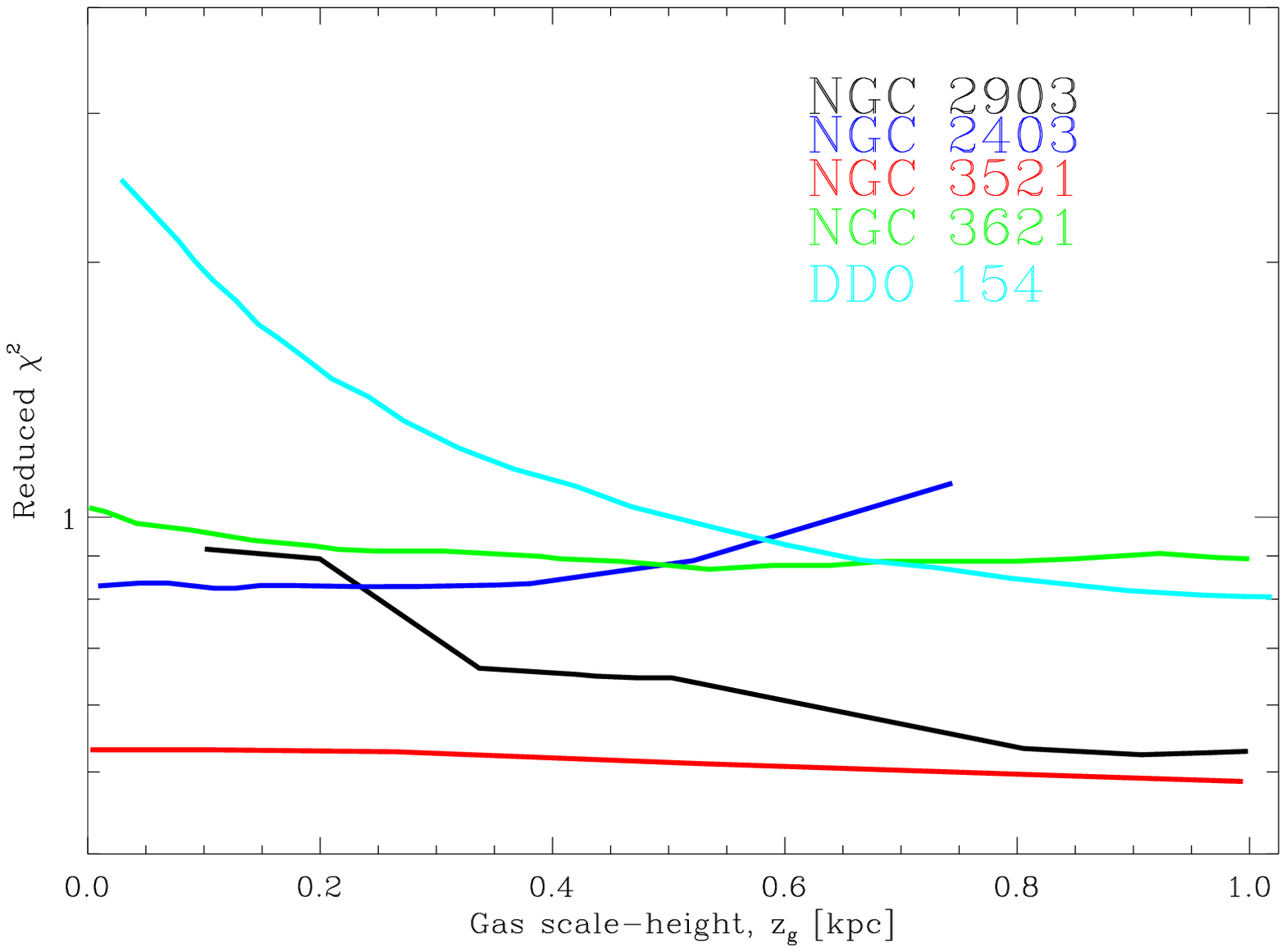}
}
\caption{Shows the curves of minimum reduced $\chi^2$ as a function of the fitted parameters of each galaxy. In the top left panel is the fitted distance of each galaxy normalised to the observed distance, and the top right is the fitted $M/L$ normalised to the $M/L$ using the diet Salpeter IMF (the figure below this is the same as above but with a larger plotted range). We also plot the two scale-heights. A different coloured line is used for each galaxy and a legend emphasises this in each panel.}
\label{fig:chi}
\end{figure*}

\protect\label{sec:res}
In Fig~\ref{fig:chi} we plot the minimum reduced $\chi^2$ for each free parameter as a function of that parameter. Specifically, this is the minimum $\chi^2$ achievable with that parameter fixed and every other parameter free. The distances and $M/L$ are limited according to the plotted range for each galaxy.

Each galaxy in Fig~\ref{fig:chi} has a different line colour. In the top left hand panel, we plot reduced $\chi^2$ for each galaxy against distance. The fitted distance is normalised to the measured distance and similarly the fitted $M/L$ is normalised to the $M/L$ predicted by the diet-Salpeter IMF. The $1\sigma$ error on distance varies from galaxy to galaxy and can be found in Table~\ref{tab:par}. For $M/L$, the Kroupa IMF is typically 70\% of the diet-Salpeter value.

Clearly two galaxies (NGC~2903 and NGC~3521) prefer substantially lower distances than the measurements suggest, whereas the other three galaxies have their minima closer to the measured distance. It is worth baring in mind that NGC~2903 and NGC~3521 have the largest uncertainties on their distances, 25\% and 30\% respectively.

The $M/L$ plots (top right and middle right) have only four lines since we fix the $M/L$ of DDO 154 to the diet-Salpeter value and do not vary it, nor do we vary its stellar scale-height. The concern here is that the curve for NGC~2903 rises sharply towards large $\chi^2$ when typical $M/L$ are tried.

The gas scale-height is only significant for two galaxies. Obviously DDO 154 is our only gas dominated galaxy and 95\% of its mass is gaseous. It prefers as large a gas scale-height as possible and the $\chi^2$ rises sharply below 0.6~kpc. On the other hand, NGC~2403 prefers a thin gas disk below 0.4~kpc in scale-height. The other three galaxies prefer gas scale-heights larger than 0.5~kpc, but no significant gain is achieved for larger values than this.

The stellar scale-height is strongly constrained by the rotation curves. All four stellar dominated galaxies show a clear minimum between 0.25 and 0.55~kpc and the quality of the fits would be significantly reduced if they were forced to be razor thin or 1~kpc in extent.

\begin{figure*}
\centering
\subfigure{
\includegraphics[angle=0,width=8.50cm]{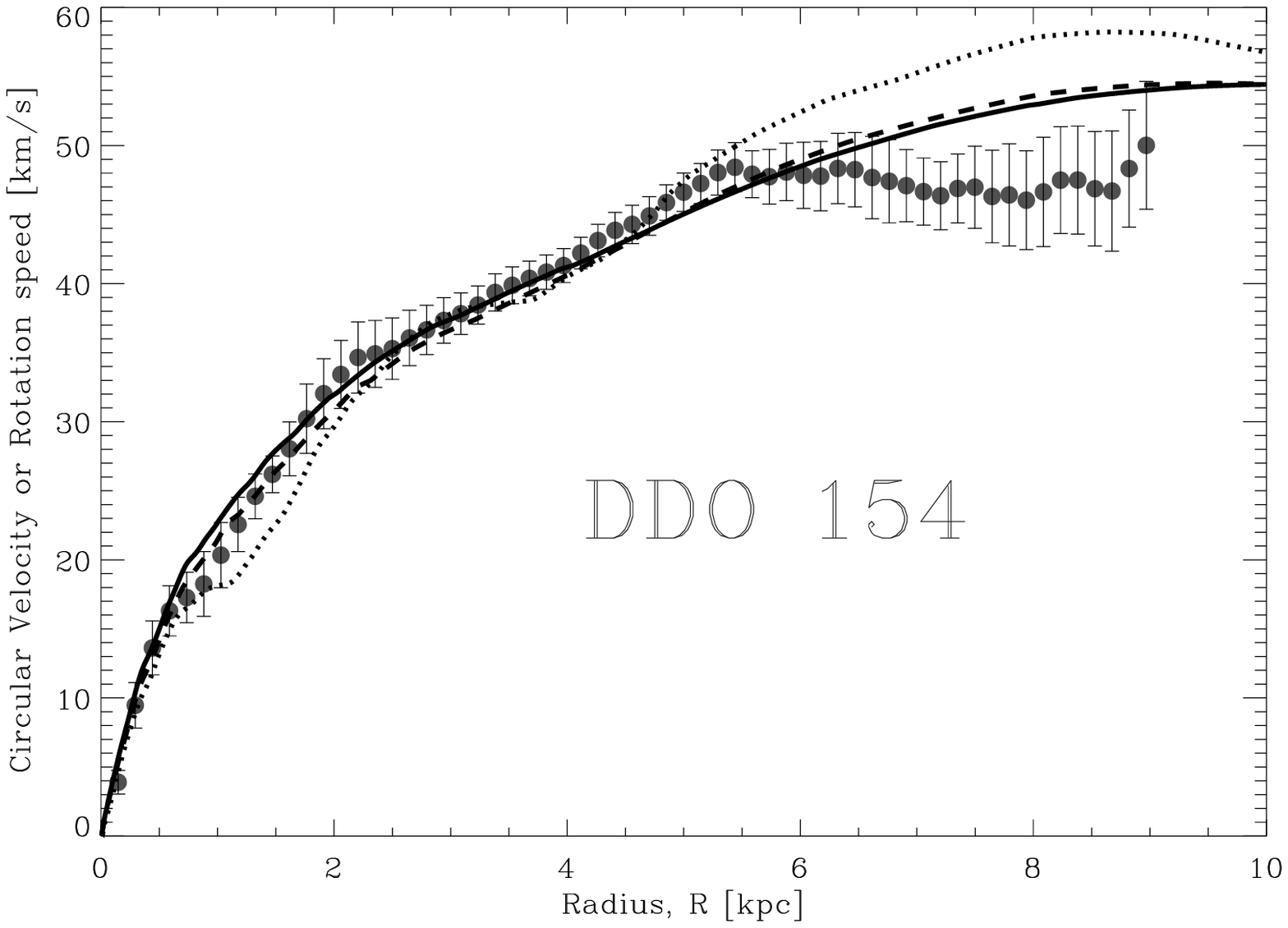}
}
\subfigure{
\includegraphics[angle=0,width=8.50cm]{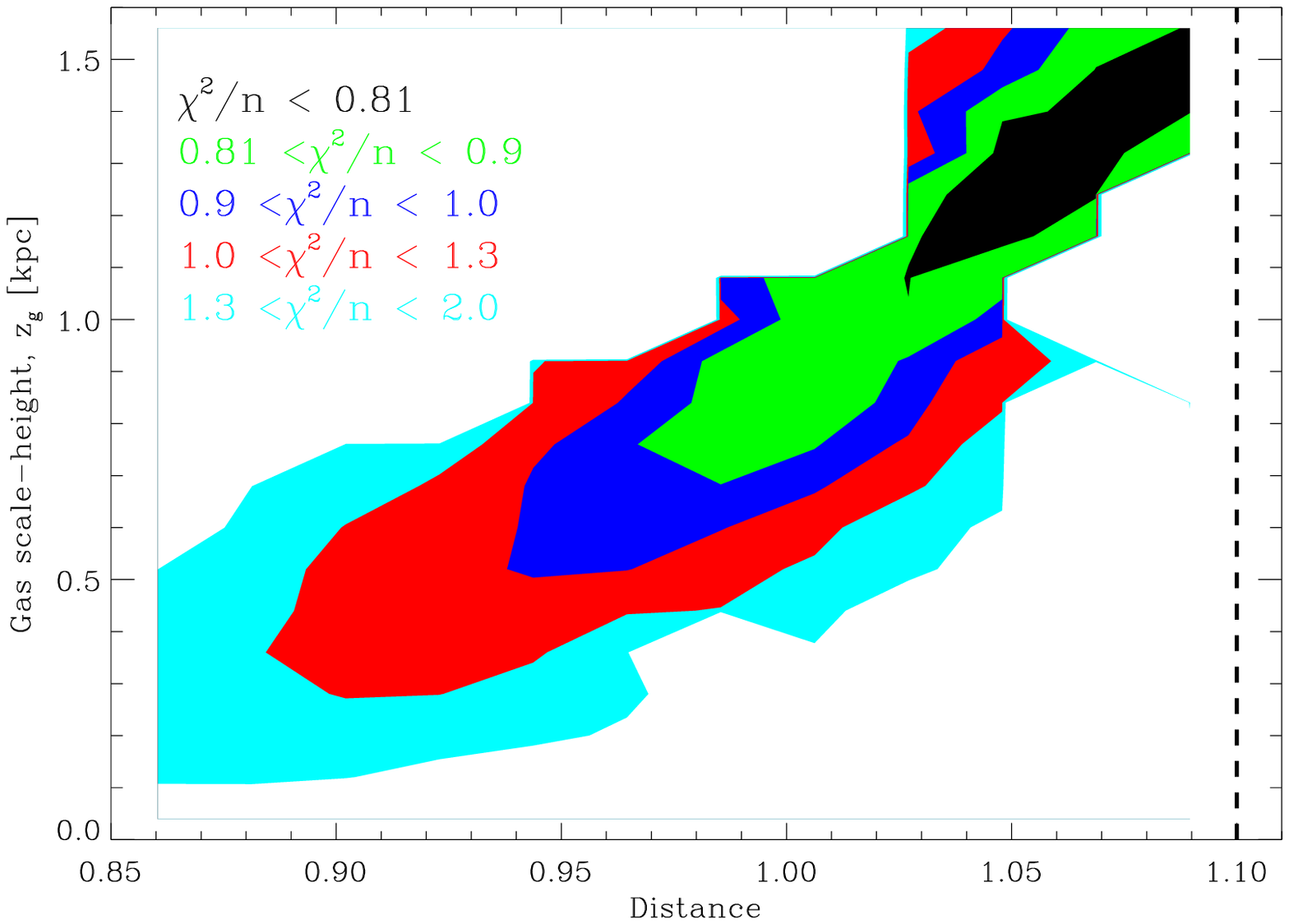}
}
\caption{In the left hand panel we plot the rotation curve of DDO~154 as measured by de Blok et al. (2008) along with our MOND fits. The curves are described in table~\ref{tab:par}, giving their line-types and model parameters. The right hand panel shows contours of reduced $\chi^2$ for the two free parameters: the distance normalised by the mean measured distance and the gas scale-height. The contour levels are defined in the panel and the vertical dashed line defines the 1$\sigma$ error on the measured distance.}
\label{fig:com154}
\end{figure*}

\begin{figure*}
\centering
\subfigure{
\includegraphics[angle=0,width=8.50cm]{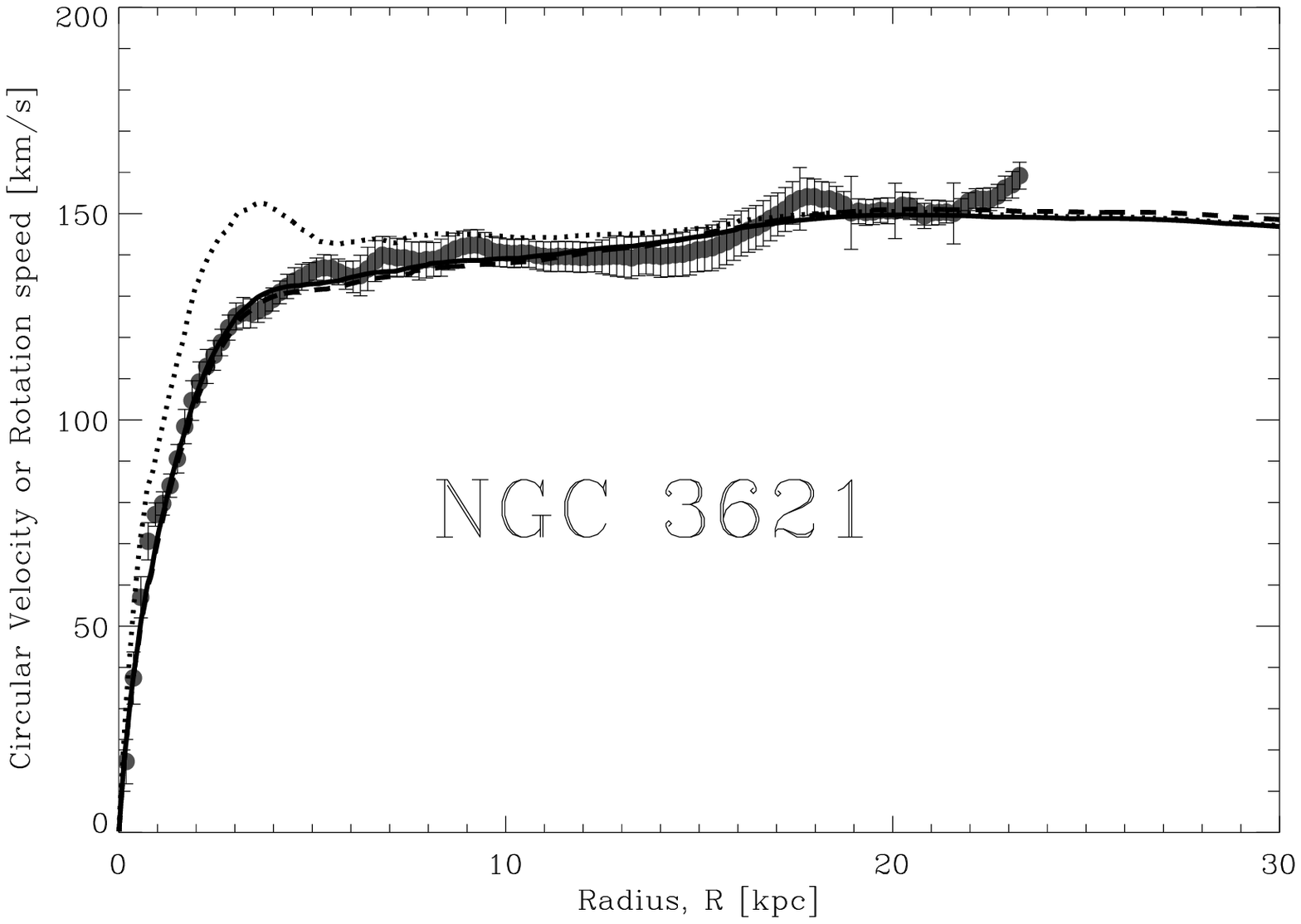}
}
\subfigure{
\includegraphics[angle=0,width=8.50cm]{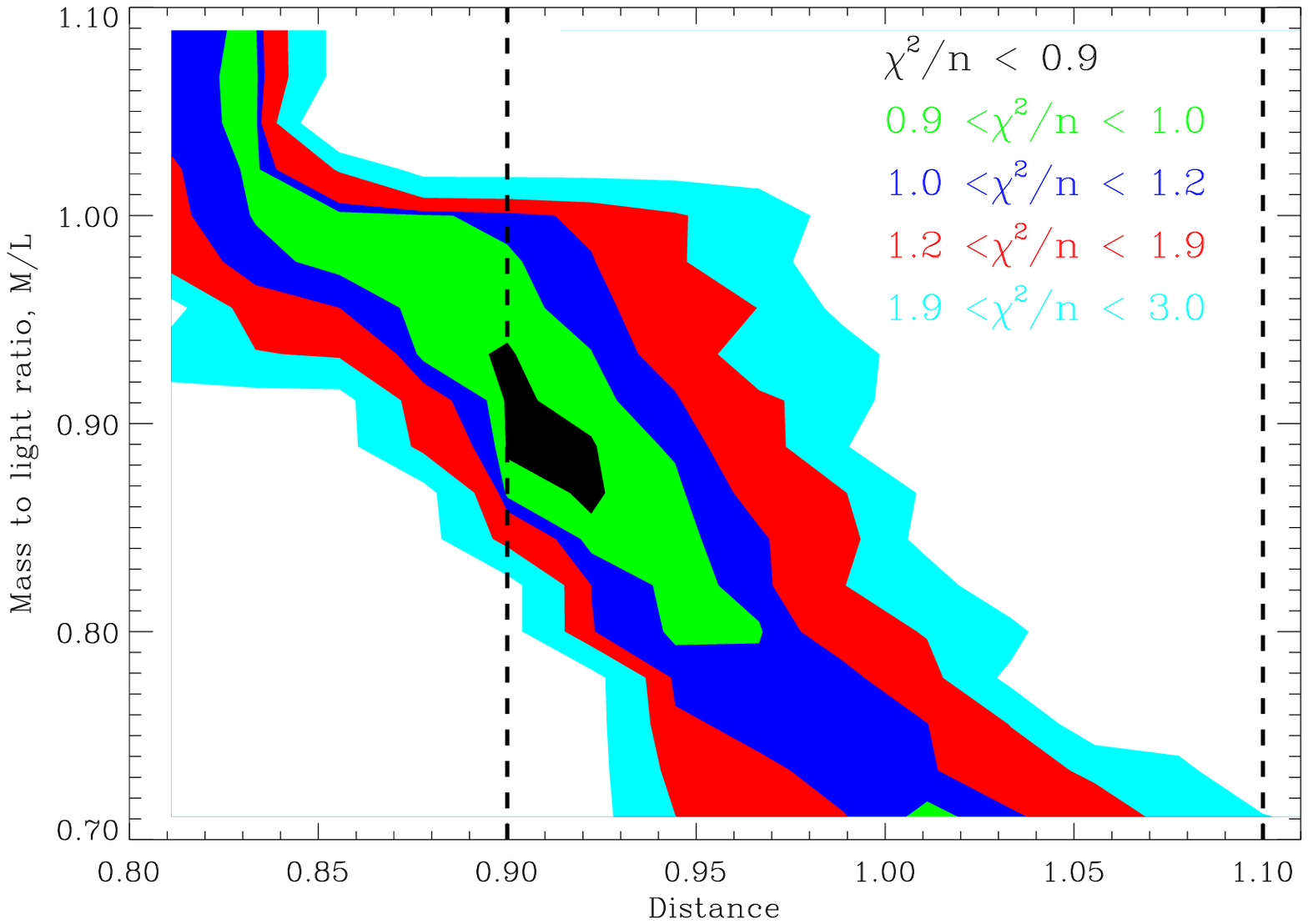}
}\\
\subfigure{
\includegraphics[angle=0,width=8.50cm]{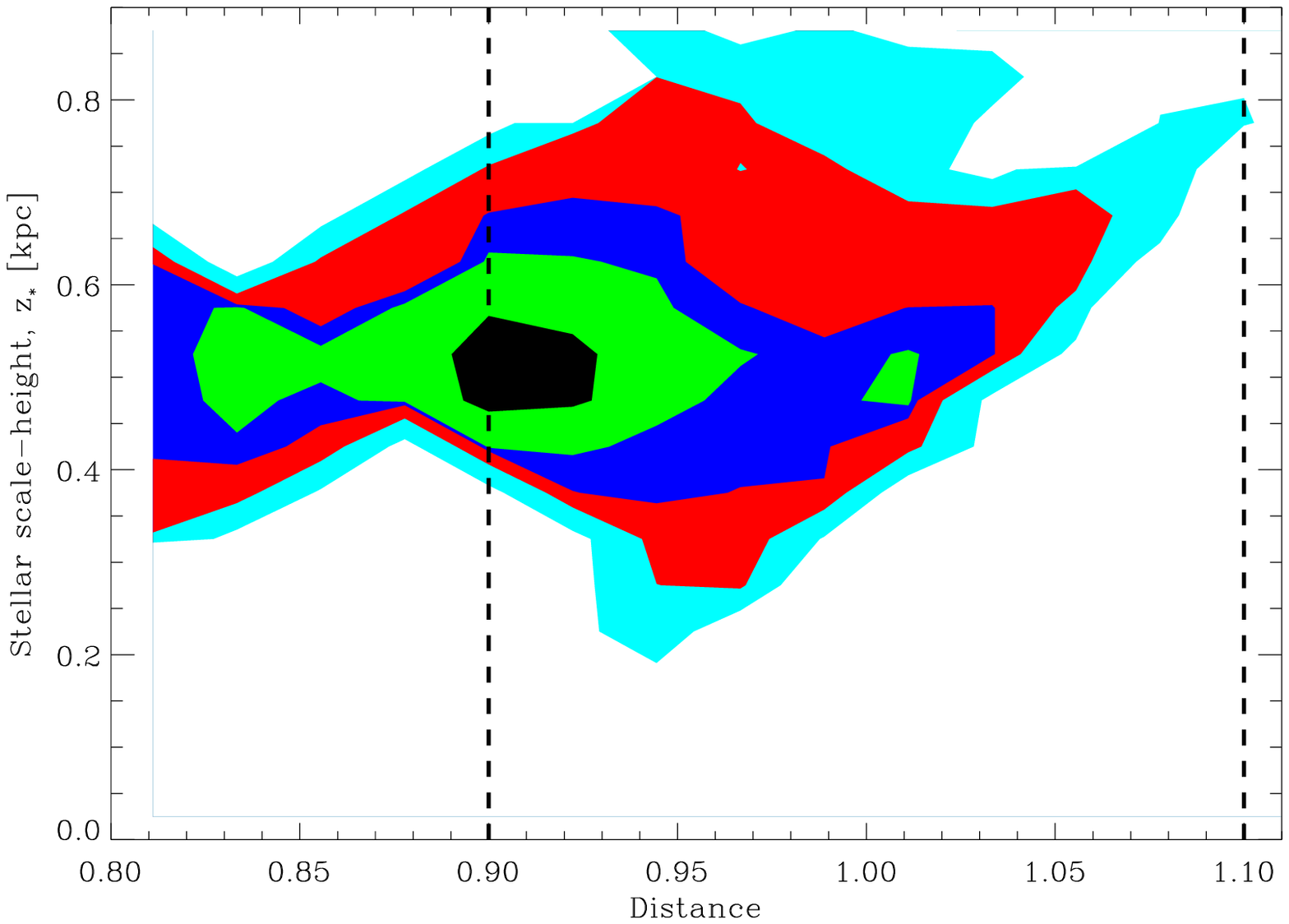}
}
\subfigure{
\includegraphics[angle=0,width=8.50cm]{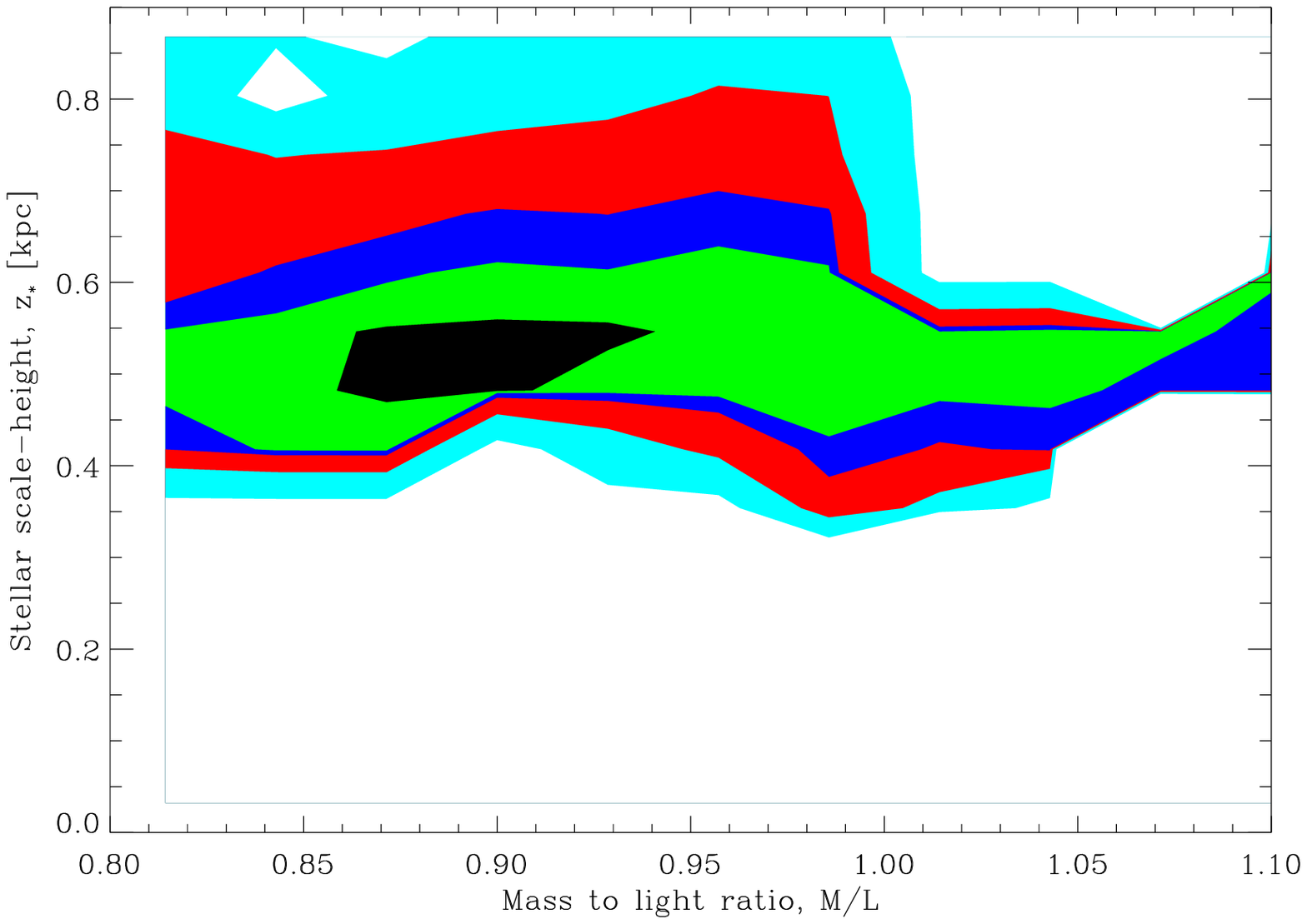}
}
\\
\caption{In the top left hand panel we plot the rotation curve of NGC~3621 as measured by de Blok et al. (2008) along with our MOND fits. The curves are described in table~\ref{tab:par}, giving their line-types and model parameters. The other three panels show contours of reduced $\chi^2$ for the three combinations of three free parameters, these are the fitted distance and $M/L$ normalised to the measured distance and diet-Salpeter IMF respectively and the fitted stellar scale height. The contour levels are defined in the top right panel. The vertical dashed line in the contour plots with distance defines the 1$\sigma$ error on the measured distance.}
\label{fig:com3621}
\end{figure*}

\begin{figure*}
\centering
\subfigure{
\includegraphics[angle=0,width=8.50cm]{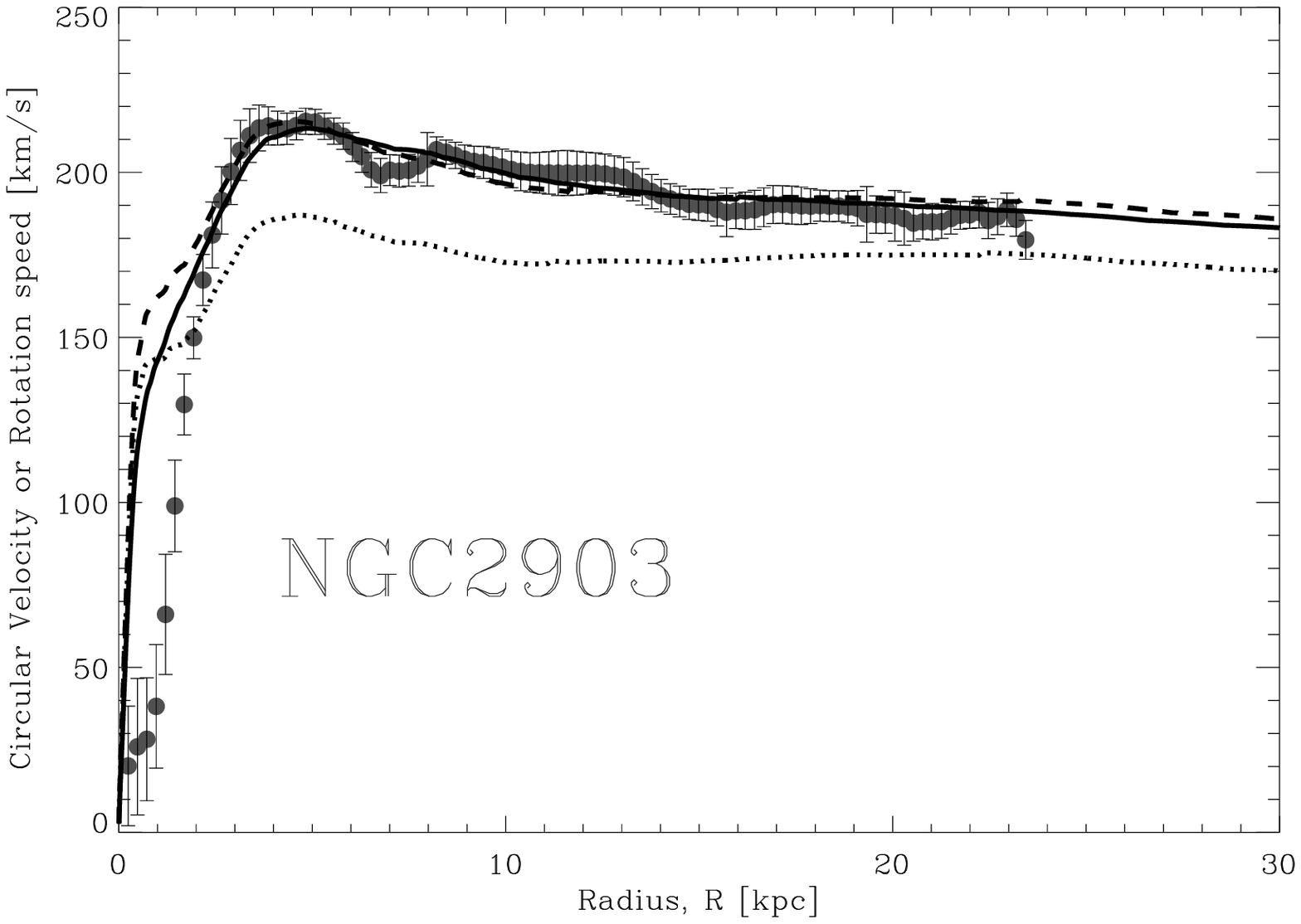}}
\subfigure{
\includegraphics[angle=0,width=8.50cm]{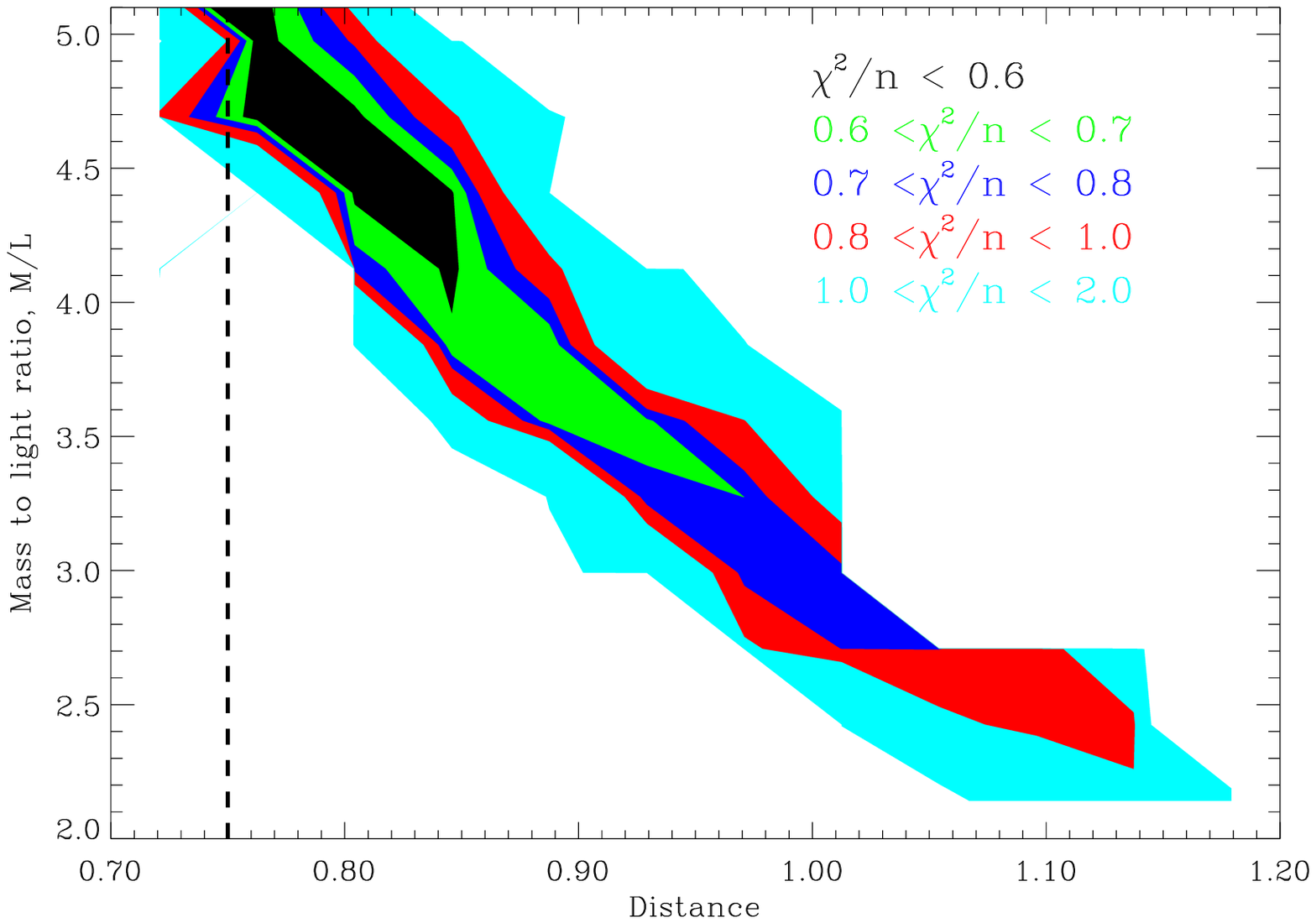}
}\\
\subfigure{
\includegraphics[angle=0,width=8.50cm]{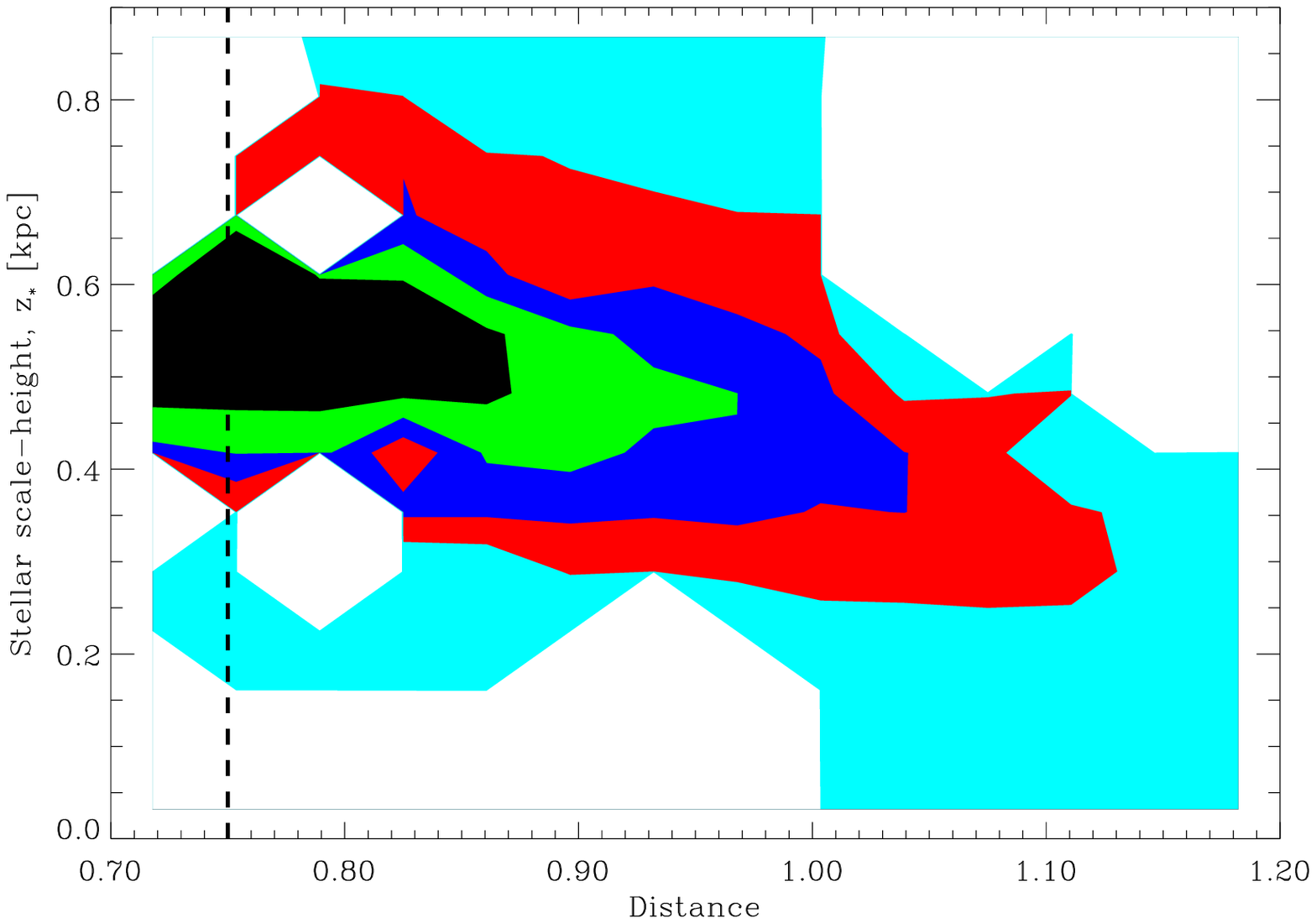}
}
\subfigure{
\includegraphics[angle=0,width=8.50cm]{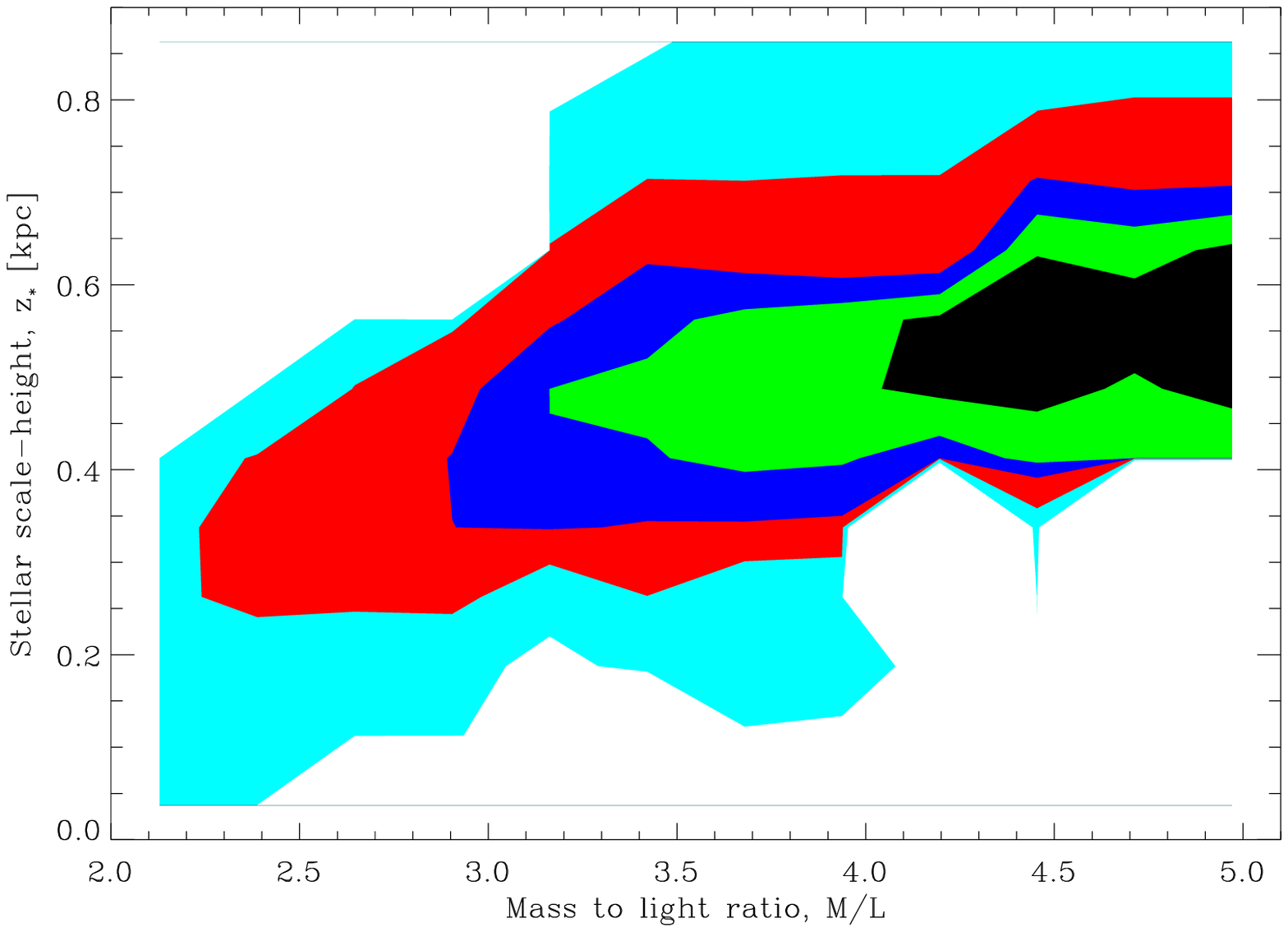}
}
\caption{As per Fig~\ref{fig:com3621}, but for NGC~2903.}
\label{fig:com2903}
\end{figure*}

\begin{figure*}
\centering
\subfigure{
\includegraphics[angle=0,width=8.50cm]{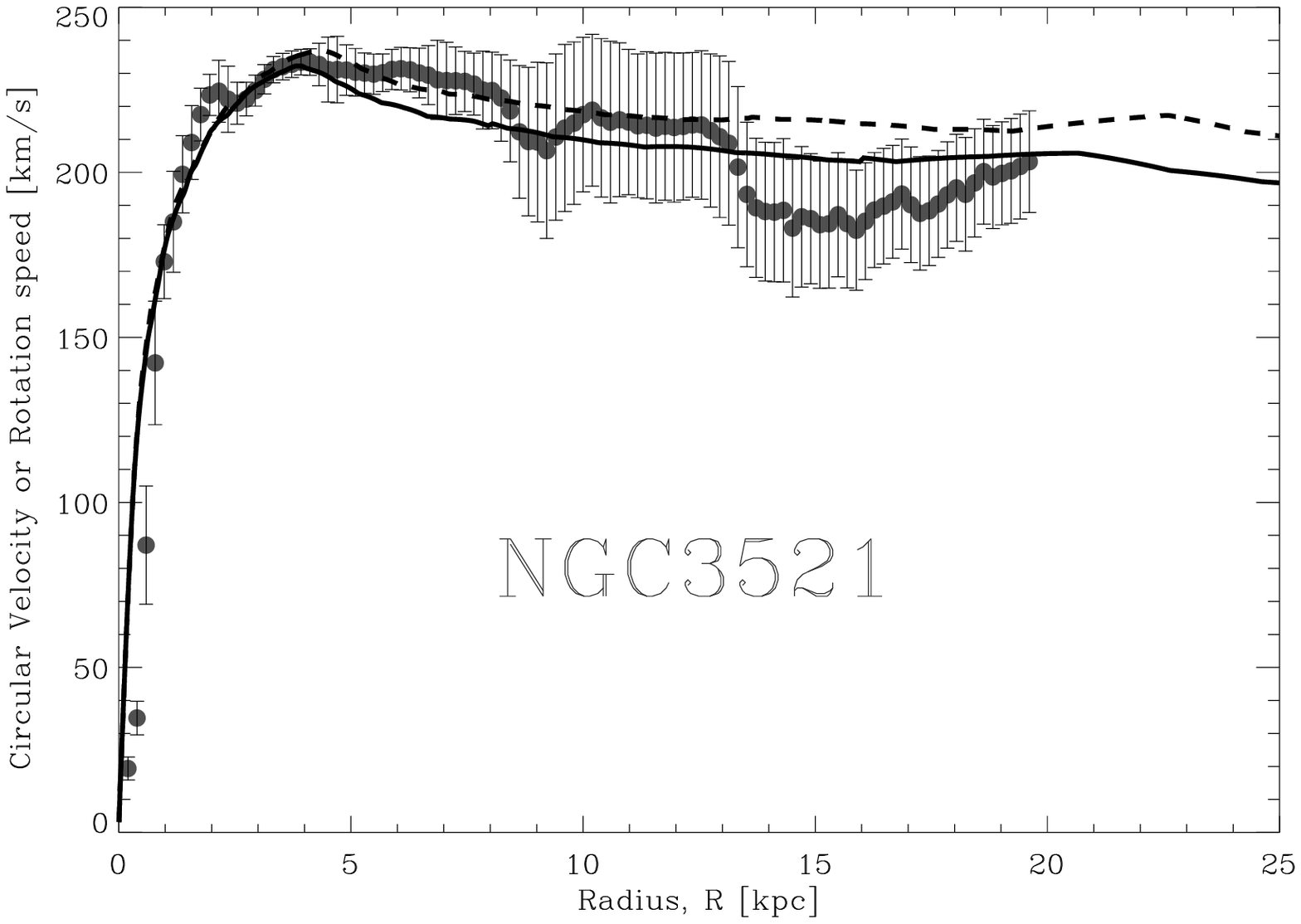}
}
\subfigure{
\includegraphics[angle=0,width=8.50cm]{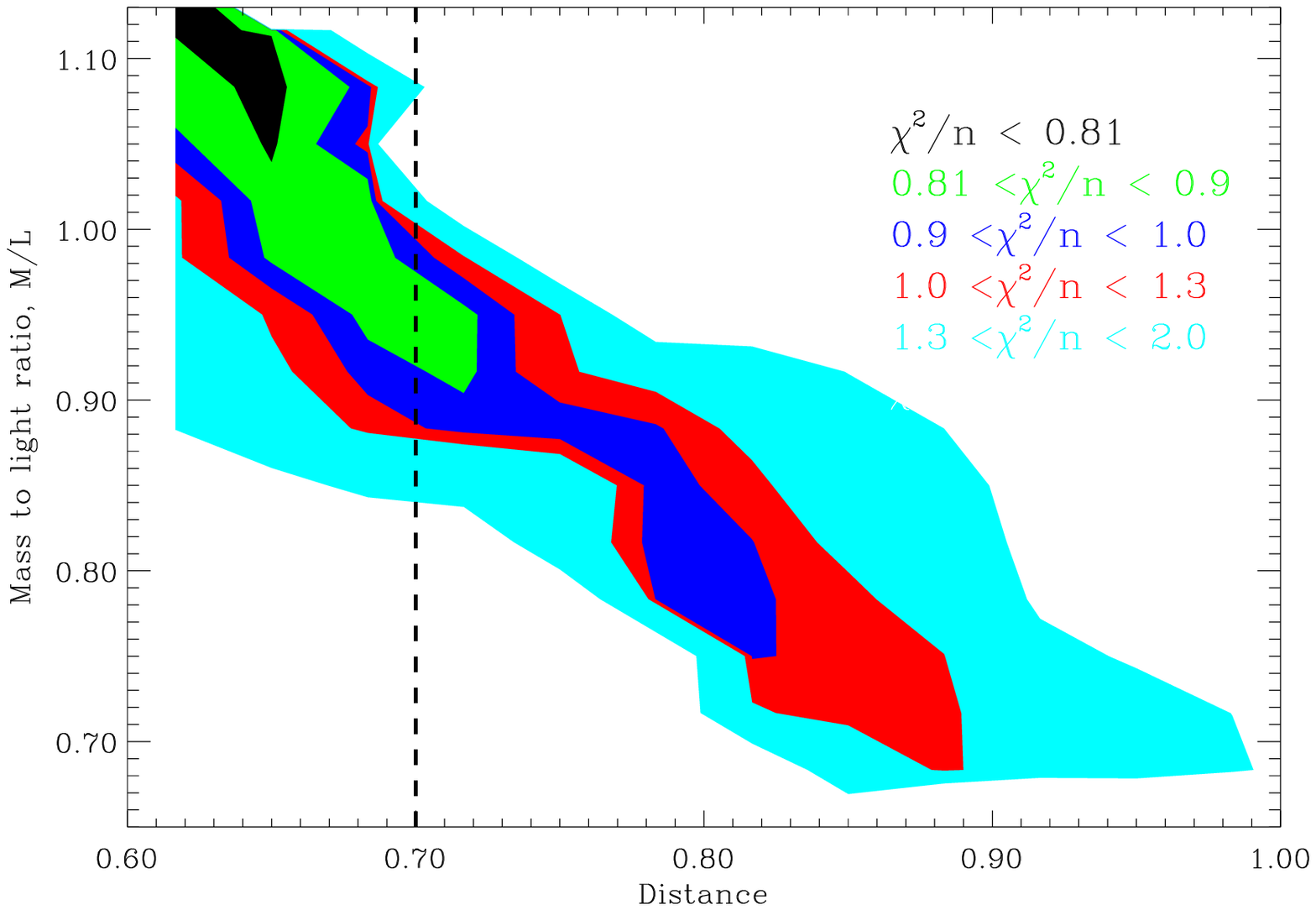}
}\\
\subfigure{
\includegraphics[angle=0,width=8.50cm]{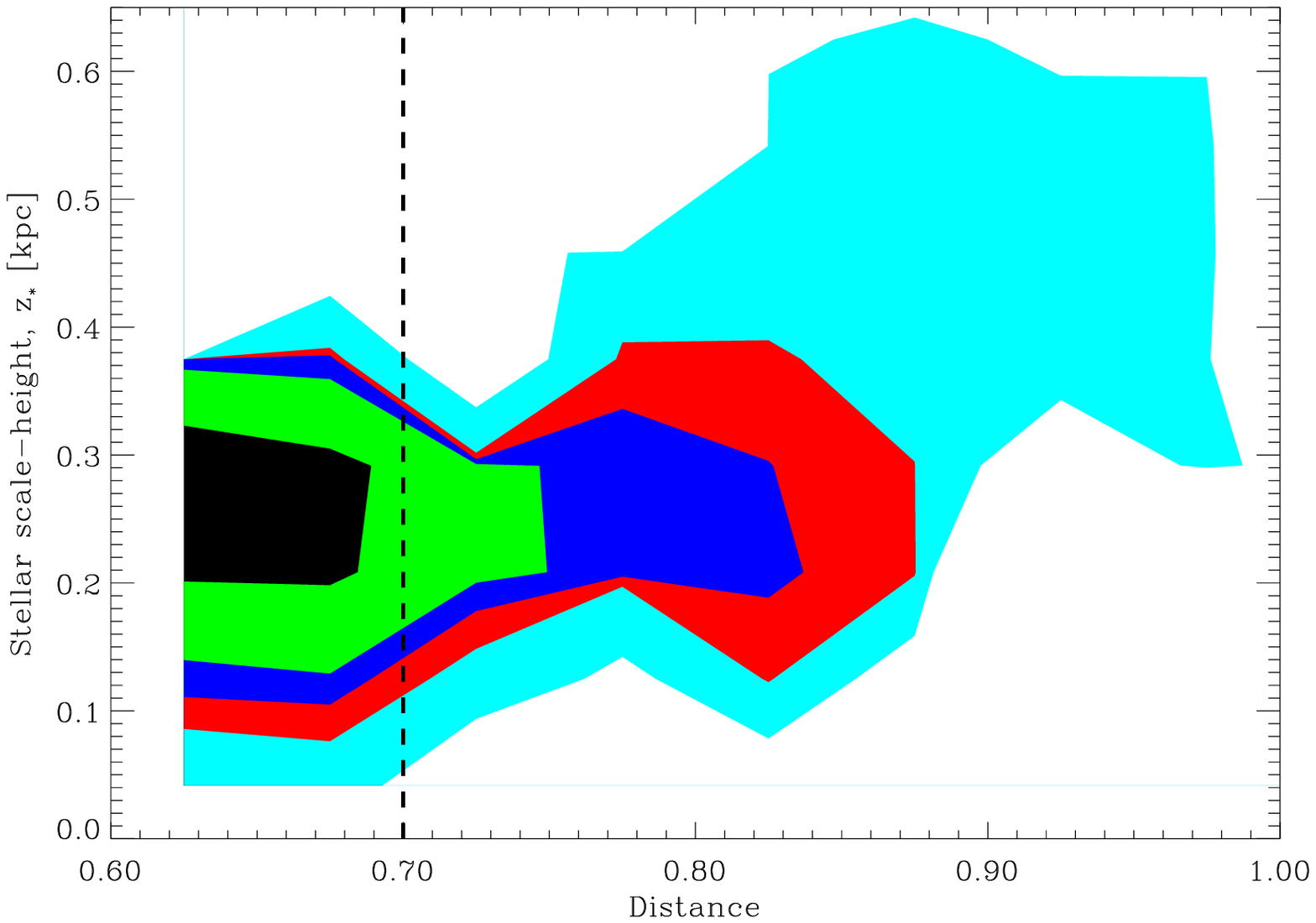}
}
\subfigure{
\includegraphics[angle=0,width=8.50cm]{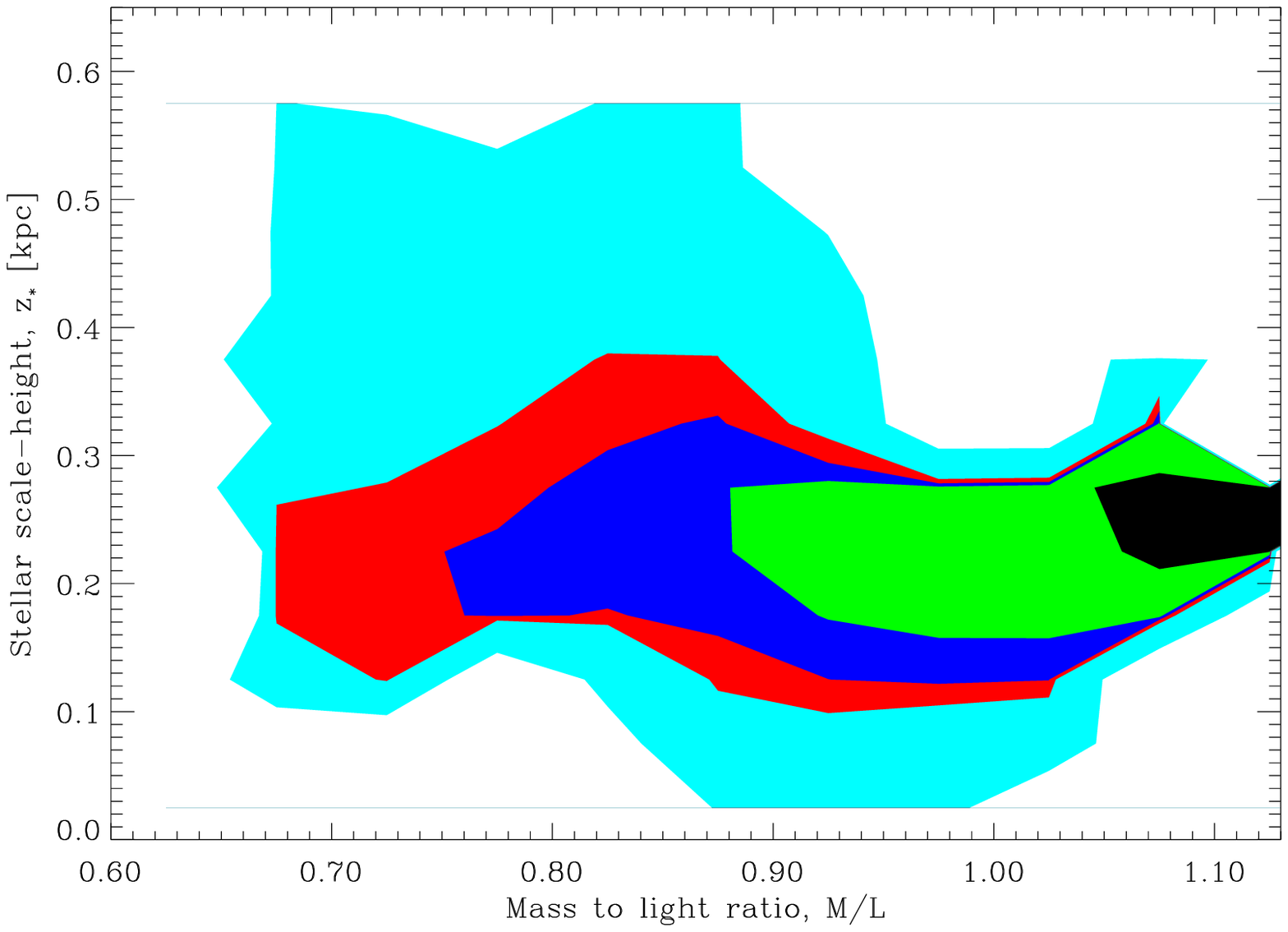}
}
\\
\caption{As per Fig~\ref{fig:com3621}, but for NGC~3521.}
\label{fig:com3521}
\end{figure*}

In Figs \ref{fig:com154}-\ref{fig:com3521} we plot various fits to the five rotation curves as well as three contour plots (DDO 154 only has one) for the three combinations of the three free parameters (gas scale-height has been omitted).

\subsection{DDO 154}
A lower $\chi^2$ can be achieved by increasing the distance and gas scale-height. In table \ref{tab:par} we give the parameters used for the three fits in the left hand panel of Fig~\ref{fig:com154}. In all figures showing the rotation curve, the fits and data points are rescaled to the distance of model (a) for that particular galaxy. In the case of DDO~154 this is 1.085 times the measured distance of 4.3~Mpc and as such is 4.67~Mpc. It is not possible to fit the inner curve and outer flat data points beyond 7~kpc with the same model. The flat outer points may be due to a warp in the outer part of its gaseous disk (\citealt{carpur98}). Nevertheless, the solid and dashed lines achieve excellent fits up to 6~kpc and the dashed line uses an acceptable distance and gas scale-height of only 0.65~kpc. The dotted line uses a razor thin gas disk and is a poor fit to the inner curve and poorer still to the outer curve. \cite{walter99} claim that dwarf galaxies (such as DDO~154) have thicker H I disks than more massive spiral galaxies due to a lower restoring force from weaker gravity. In order to be a good fit in MOND, the outer warp should be responsible for the anomalously low rotation speeds between 7 and 8.5~kpc.

\subsection{NGC 3621}
The $\chi^2$ minimum for NGC~3621 is obvious in the contour plots of Fig~\ref{fig:com3621} and puts strong constraints on all three parameters. The fits, whose parameters are given in table \ref{tab:par}, at the top left hand panel show the best fit (solid line) and another fit with slightly larger distance (dashed line), more in keeping with the measured distance. Both are excellent fits. The dotted line shows the best fit model when the best fit stellar scale-height is replaced with a razor thin stellar disk and is a very poor fit to the central rotation curve. The best fit parameters are all within acceptable bounds and this is an excellent fit for MOND.

\subsection{NGC 2903}
The contour plots in Fig~\ref{fig:com2903} show that in order to have a satisfactory fit, regardless of distance and stellar scale-height, the $M/L$ must be more than twice the predicted value. If this is the case, then an excellent fit can be achieved as shown in the top left panel. On the other hand, using a $M/L$ of only 50\% larger than the predicted value gives a poor fit, as the dotted line attests. Therefore, NGC~2903 is a problem for MOND unless a reason can be found why it should have a $M/L$ that is twice the diet-Salpeter prediction. If such a reason was found, it would be well fit by MOND.

\subsection{NGC 2403}
\begin{figure*}
\centering
\subfigure{
\includegraphics[angle=0,width=8.50cm]{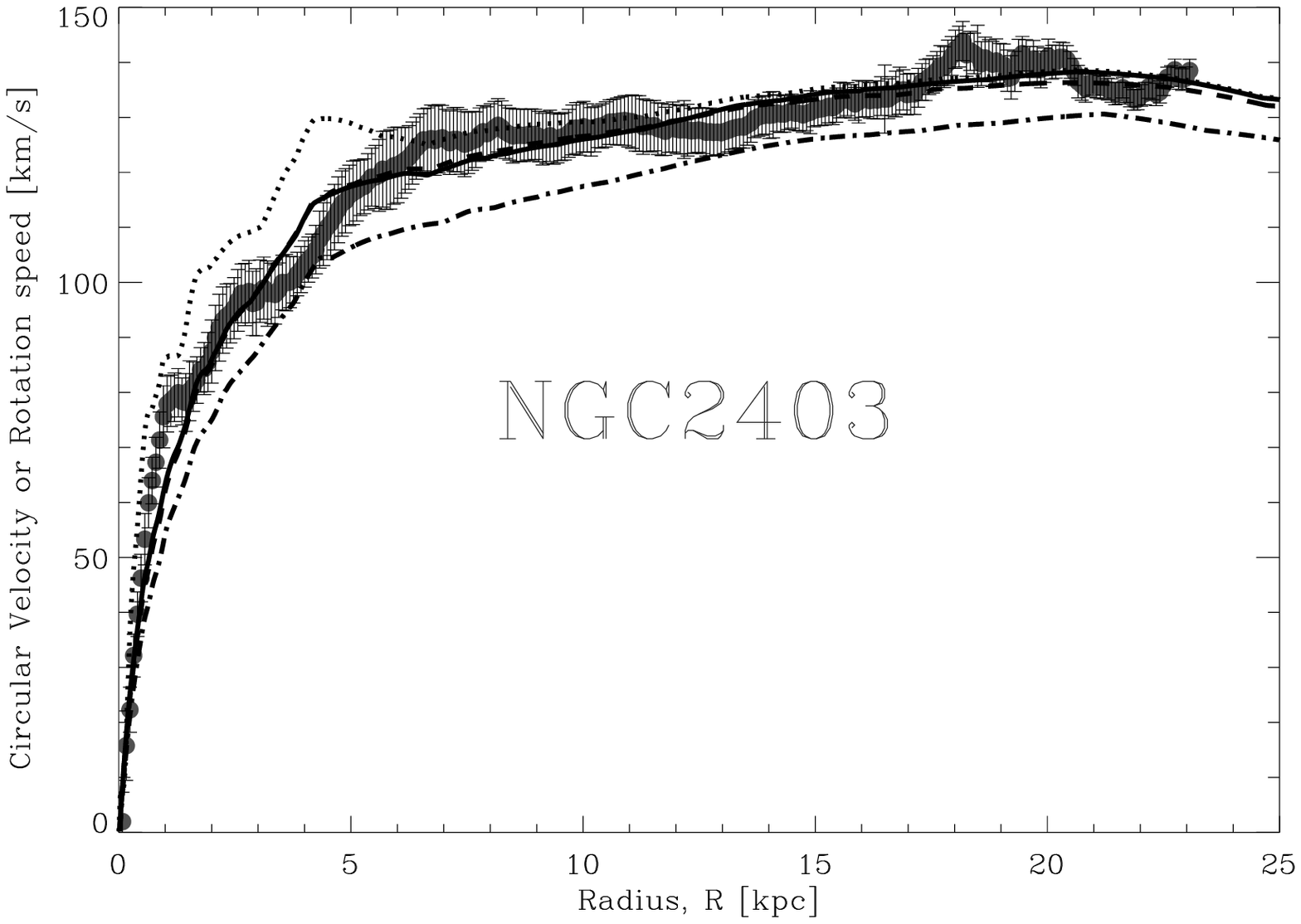}
}
\subfigure{
\includegraphics[angle=0,width=8.50cm]{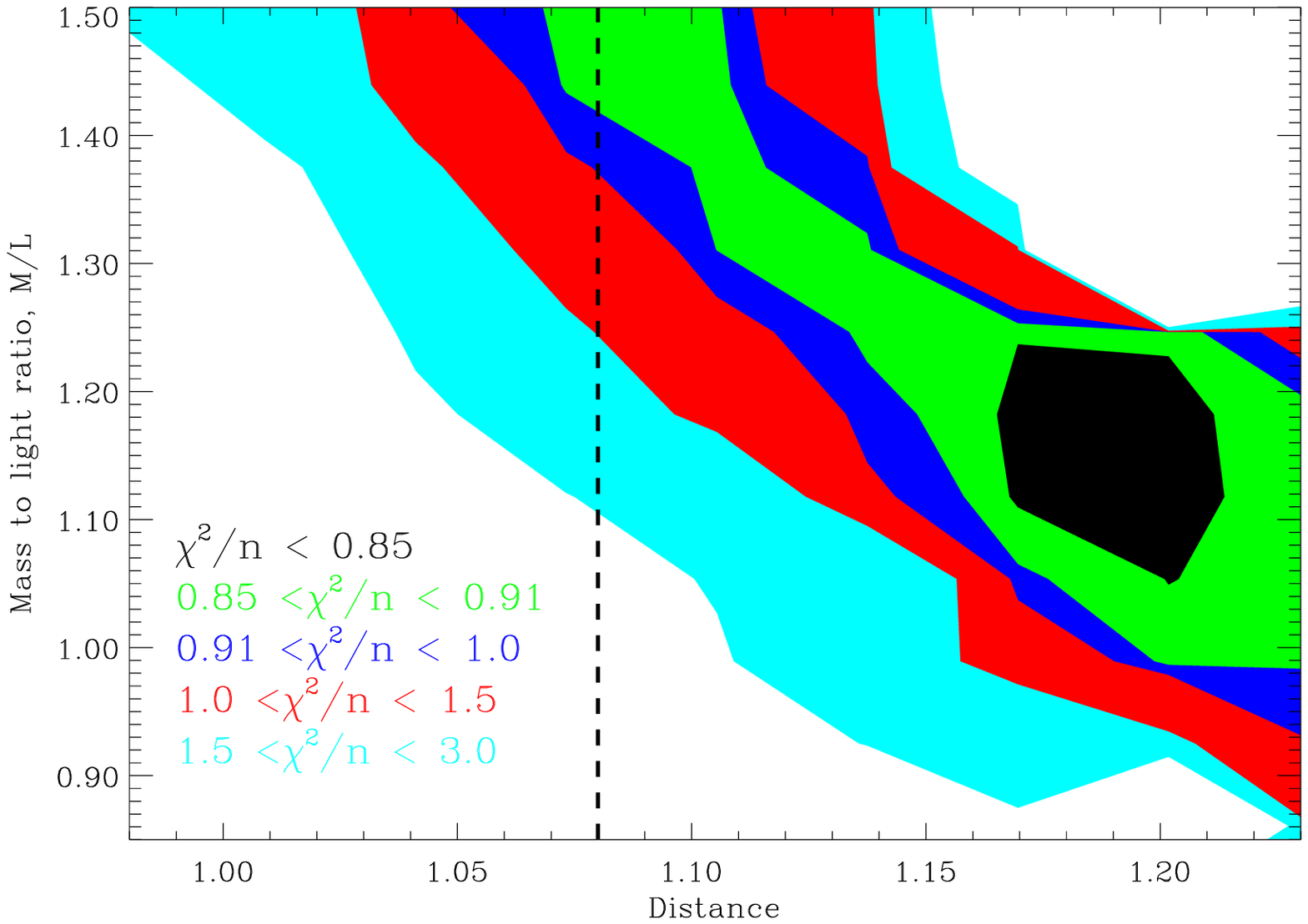}
}\\
\subfigure{
\includegraphics[angle=0,width=8.50cm]{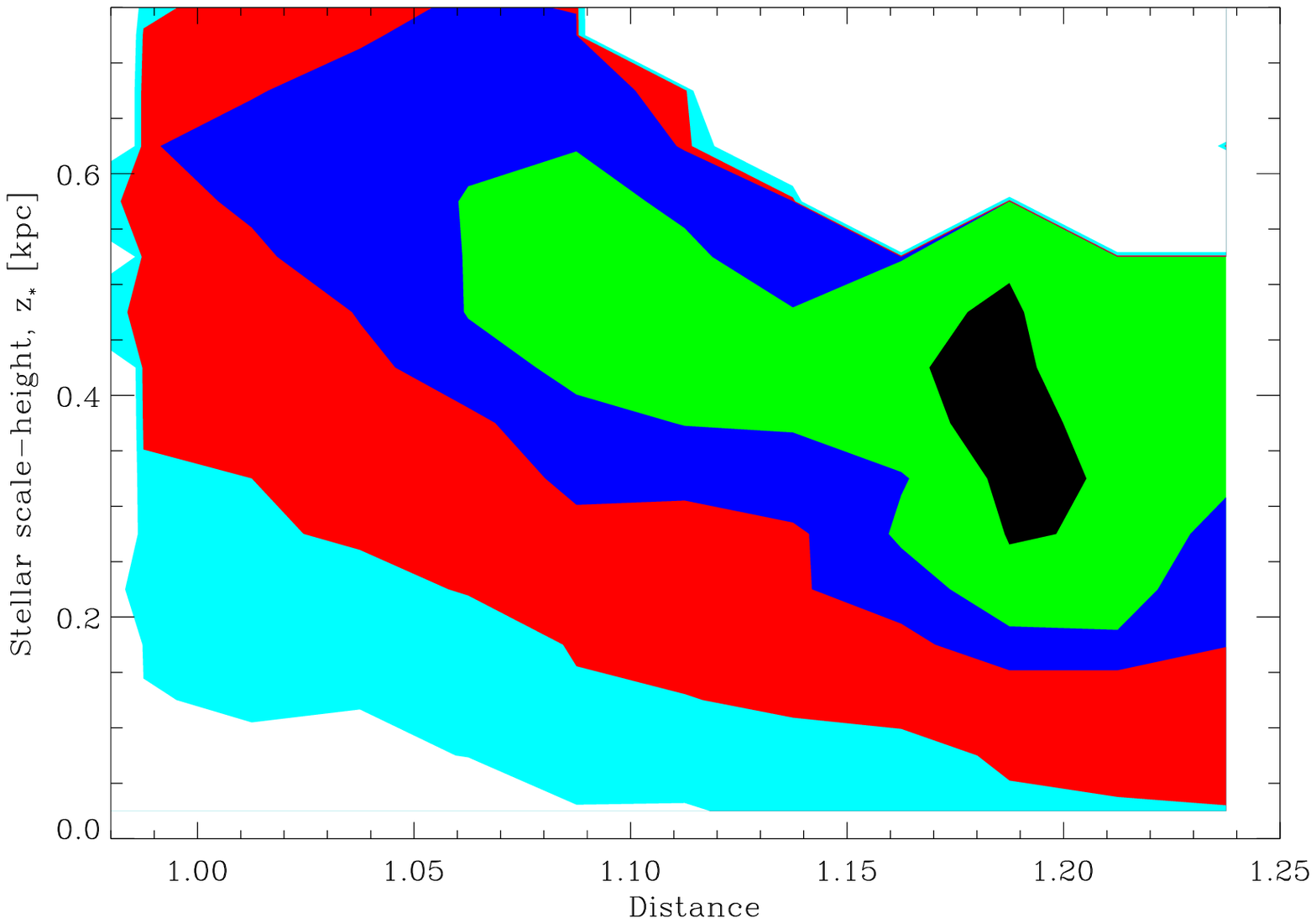}
}
\subfigure{
\includegraphics[angle=0,width=8.50cm]{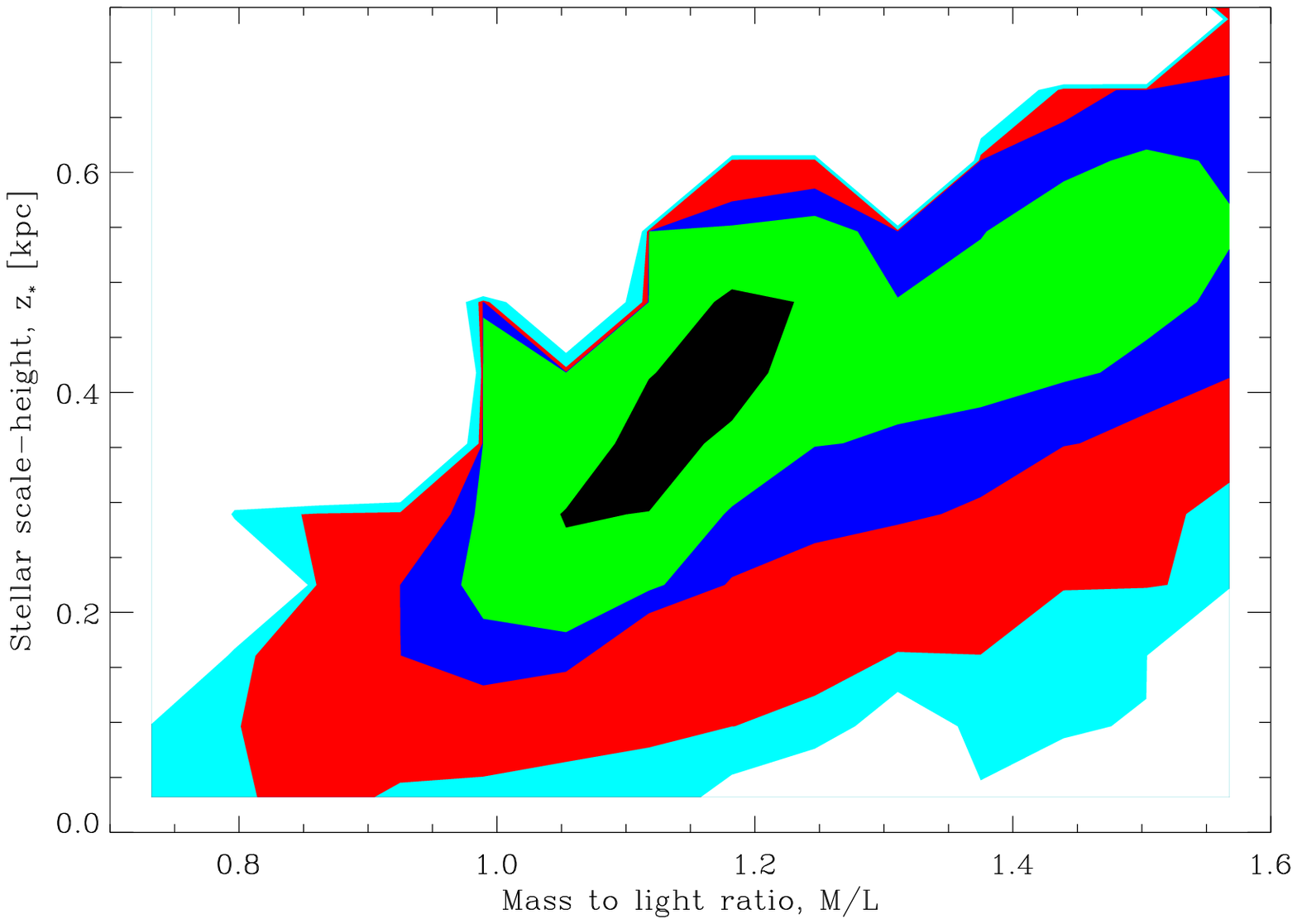}
}
\caption{As per Fig~\ref{fig:com3621}, but for NGC~2403.}
\label{fig:com2403}
\end{figure*}

In Fig~\ref{fig:com2403} the contour plot of distance against $M/L$ shows the requirement for a good fit to have either an excessively large $M/L$ or distance. This is aggravated by the tight error on the distance of a mere 8\% and the relatively low predicted $M/L$ for the given stellar population. For instance, the predicted 3.6~$\mu m$ diet-Salpeter IMF $M/L$ of NGC~2403 is 0.39, whereas for NGC~2903, 3521 and 3621 it is 0.61, 0.73 and 0.59 respectively.

In the top left hand panel of Fig~\ref{fig:com2403} we fit four curves to the observed rotation curve. The solid and dashed curves are the best fit models, which respectively use a $>2\sigma$ discrepant fitted distance and regular $M/L$, and a more regular distance but more discrepant $M/L$. The other two lines are a model with the same parameters as the best fit, but using a razor thin stellar disk (dotted line), and one with a regular distance and $M/L$ (dot-dashed line). Both of these are unsatisfactory fits. As with NGC~2903, this galaxy NGC~2403 can only be consistent with MOND if the cause of its higher than expected $M/L$ is found, however, this is a much less significant problem than NGC~2903 because even with the required $M/L$ for a good fit, it would still have a typical $M/L$ compared to other galaxies. 

\subsection{NGC 3521}
For NGC 3521, the preference is for a much lower distance than measured, which can be seen in the contour plot of Fig~\ref{fig:com3521}. This is not necessarily out of the question since the error on its distance is 30\%, but the $\chi^2$ minimum is actually further from the mean than this. The parameters of the two models plotted against the observed rotation curve in the top left panel are given in table \ref{tab:par} and correspond to a distance more than 1$\sigma$ lower than the mean and roughly $0.5\sigma$ lower than the mean. Both require sensible $M/L$ and the stellar scale-height is constrained to be close to 0.25~kpc. The fit with the less discrepant distance (dashed line) is a better fit to the inner 13~kpc and the discontinuity in measured rotation speed at this radius makes it difficult for any model to achieve a good fit to both the inner and outer curve. In general, this galaxy is well fit by MOND.

\subsection{Stellar scale-heights}
A somewhat surprising result is the tight constraint placed on the values of the stellar scale-height for the four stellar dominated galaxies and the gas scale-height of DDO 154. From the contour plots of Figs \ref{fig:com154} to \ref{fig:com3521} we see the relative lack of freedom in the stellar scale-heights and this gives us the opportunity to put a further constraint on the model by checking if the required scale-heights are consistent with the typical stellar vertical velocity dispersions of other spiral galaxies.

\begin{figure*}
\centering
\subfigure{
\includegraphics[angle=0,width=8.50cm]{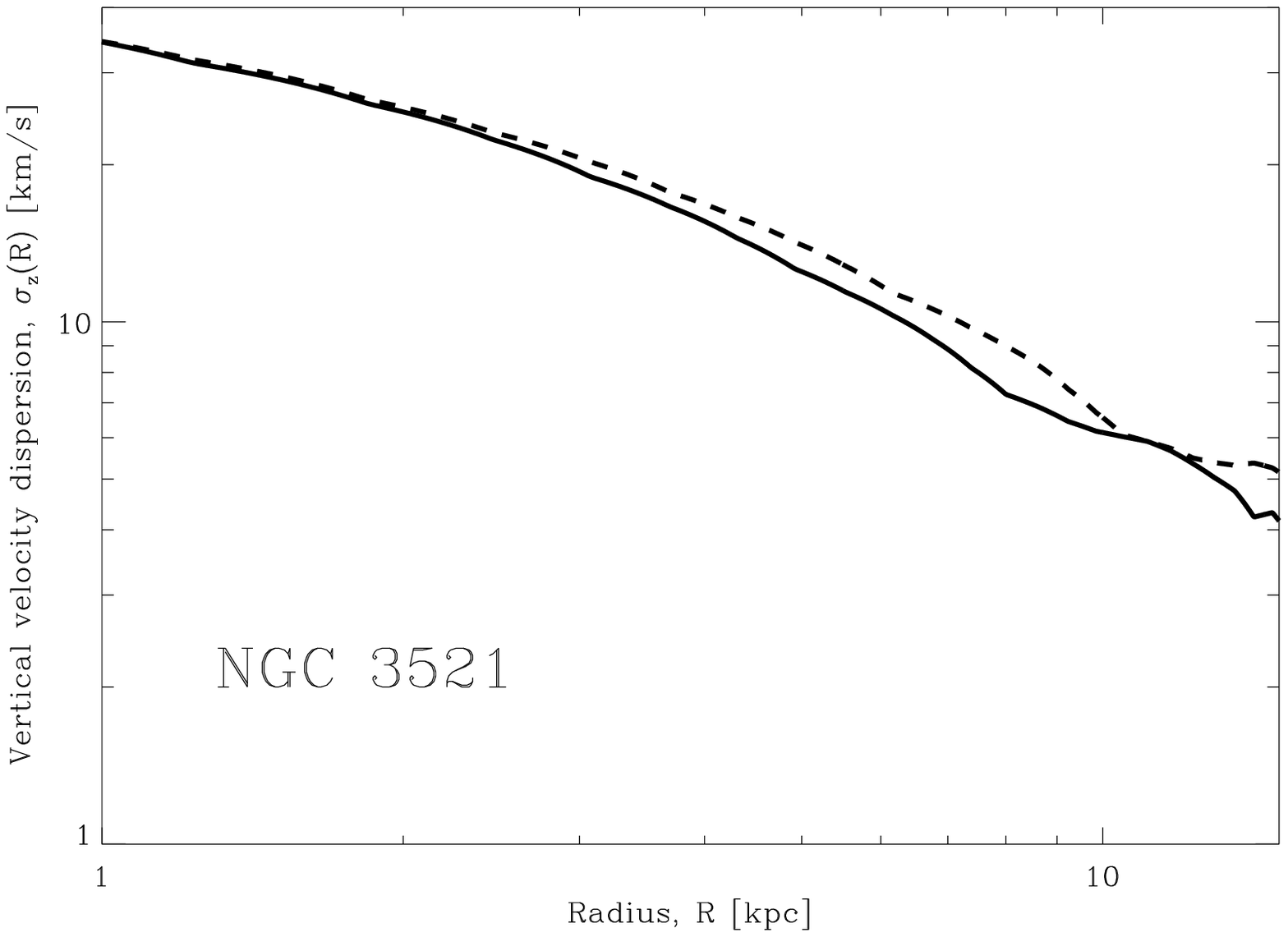}
}
\subfigure{
\includegraphics[angle=0,width=8.50cm]{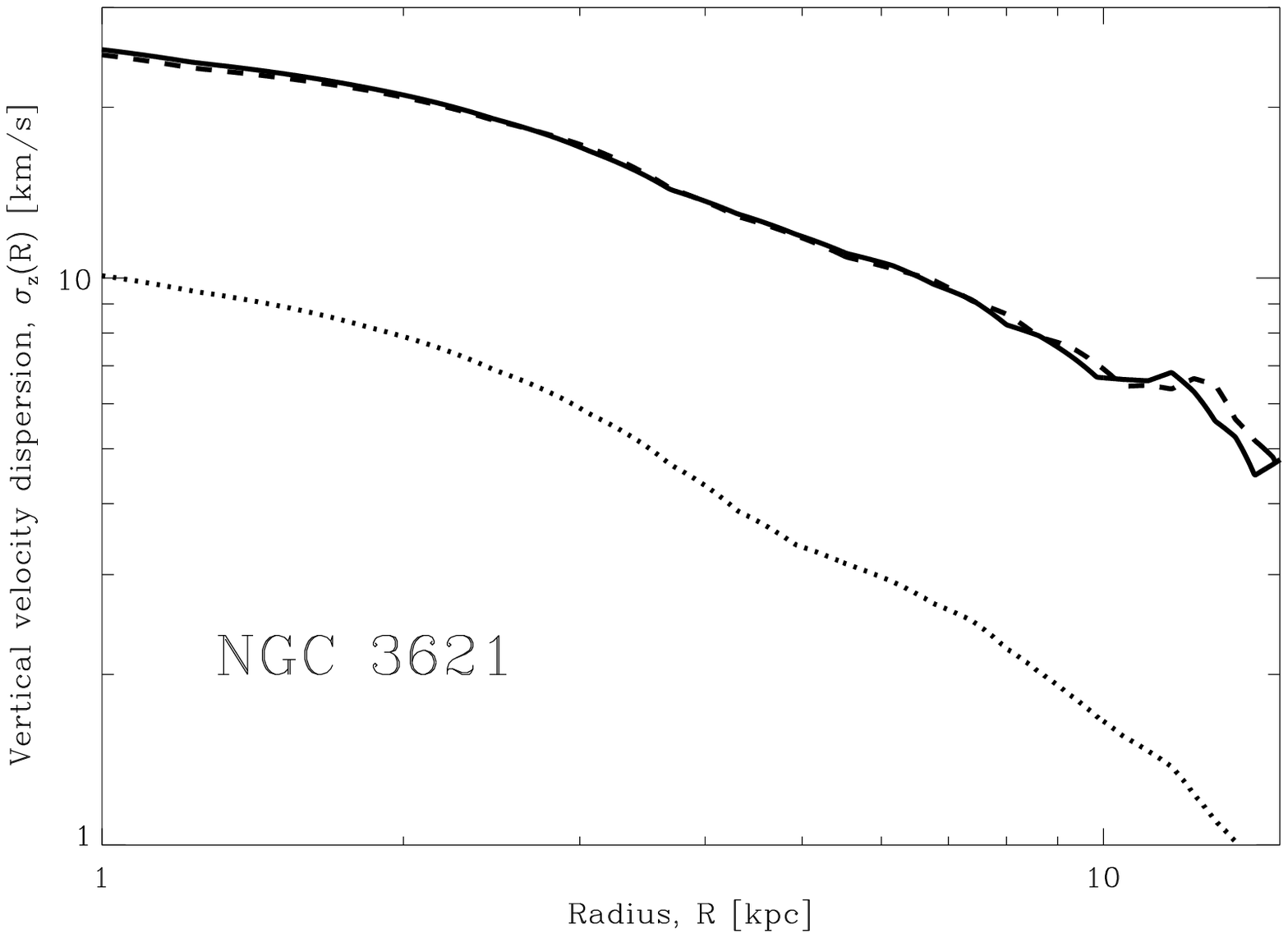}
}\\
\subfigure{
\includegraphics[angle=0,width=8.50cm]{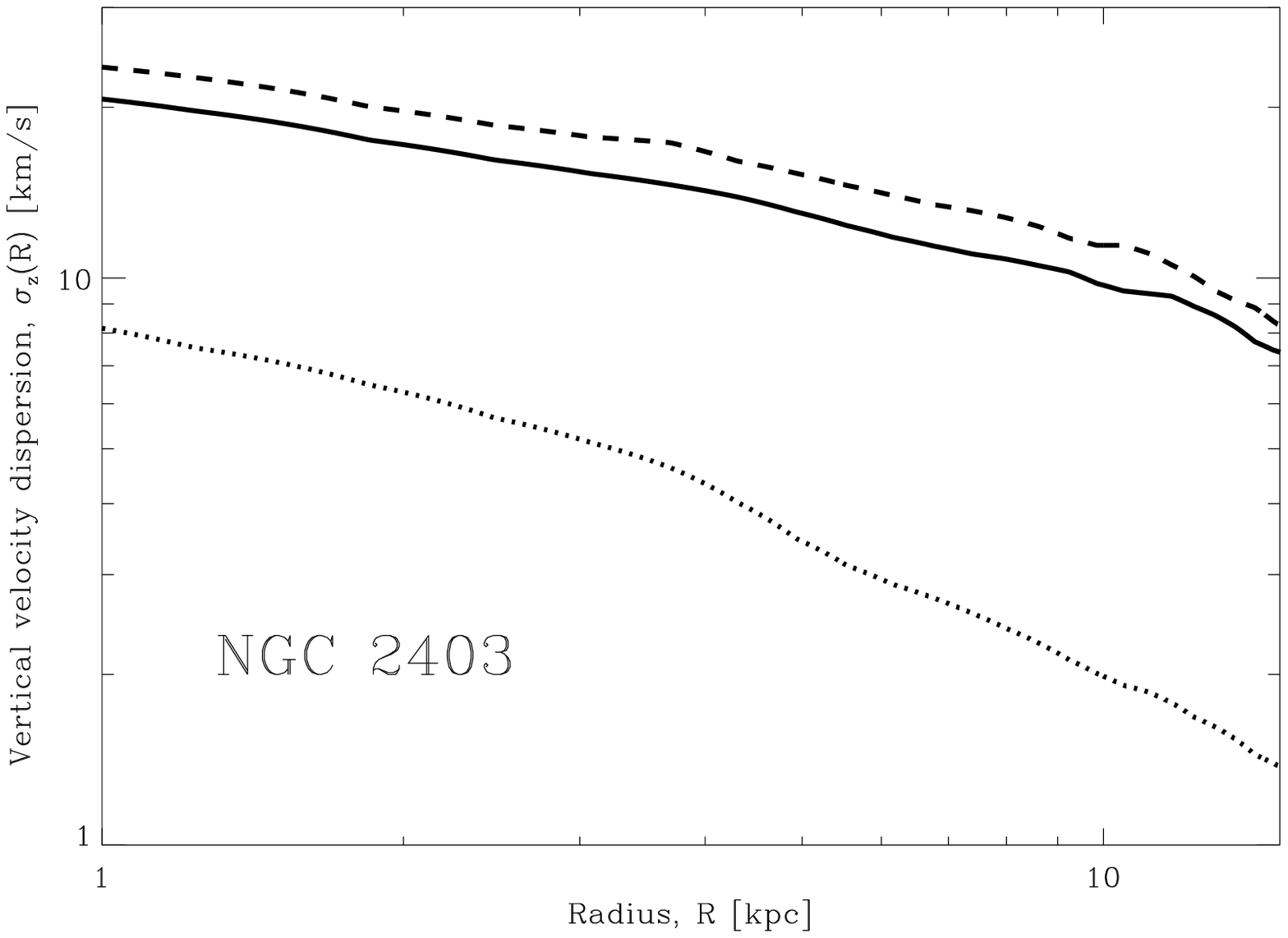}
}
\subfigure{
\includegraphics[angle=0,width=8.50cm]{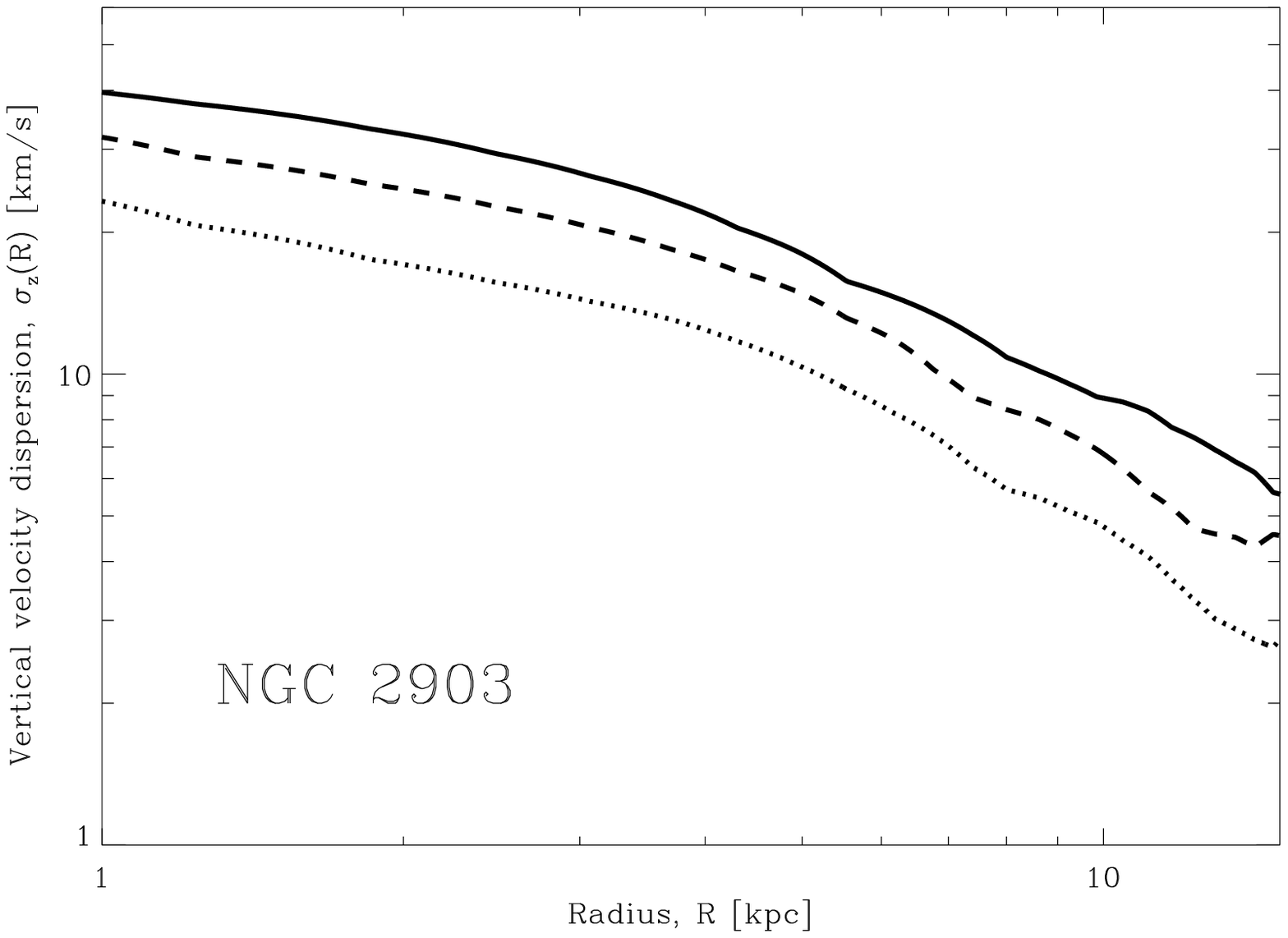}
}
\\
\caption{Stellar vertical velocity dispersions as functions of radius for the four stellar dominated galaxies. The linetypes for each galaxy are defined such that the solid, dotted and dashed lines always corresponds to models a, b and c respectively as set out in table \ref{tab:par}.}
\label{fig:scah}
\end{figure*}

Making the assumption that the stellar velocity dispersions are isothermal with height above the disk we can use the following equation to solve for the stellar vertical velocity dispersion at any radius

\beq
\sigma_{z,*}^2(R)  = -{d\Phi(R,z) \over dz} \times {\rho_*(R,z) \over \partial_z\rho_*(R,z)}.
\eeq
A similar test could be used to constrain the gas scale-heights, but we don't follow that route. The right hand side is comprised fully of quantities that are known in the rotation curve fitting code. As can be seen in Fig\ref{fig:scah}, the vertical velocity dispersions of all the galaxies have a similar shape. Generally, for the best fit models (solid and dashed lines) the trend is to have a large central vertical velocity dispersion $\sigma_z(0)$ of between 20 and 35$\kms$ and this drops to between 4 and 8$\kms$ by 15~kpc. The dotted lines, which are for very thin disks of only 0.05~kpc scale-height, have much lower velocity dispersions - with NGC 2403 and NGC 3621 being centrally just 8 and 10~$\kms$ respectively.

\cite{bottema93} studied the stellar velocity dispersions of a sample of 12 spiral galaxies of varying blue band magnitude in the range $M_B$=-18.76 to -22.22. They found a trend for the vertical stellar velocity dispersions to scale with decreasing galaxy magnitude (increasing luminosity). The three galaxies with $M_B>-20$, for which only NGC~3521 from our sample does not apply, were NGC~3198, 3938 and 6503 and they had central stellar velocity dispersions of $45\pm5~\kms$, $32\pm13~\kms$ and $30\pm7~\kms$ respectively, although NGC~3198 and NGC~3938 are inclined. The stellar velocity dispersions dropped to between 10 and 20~$\kms$ after 2 or 3 scale-heights, which is similar to our models. The larger galaxies with $M_B<-21$, which applies to NGC~3521, had central stellar velocity dispersions between 50 and 120$\kms$. This is larger than the central stellar velocity dispersion of NGC~3521. Even increasing the stellar scale-height to 1~kpc can only increase the central vertical stellar velocity dispersion to $60\kms$ and this remains on the low end of the sample studied by \cite{bottema93}. Perhaps the increased velocity dispersions with luminosity are also linked to the increased prevalence of bulges with luminosity and NGC~3521's lack of a significant spheroid bucks this trend.

The stellar and gas scale-heights we use in the fits are given in Table~\ref{tab:par} along with the stellar scale-heights suggested by \cite{deblok08}. The suggested stellar scale-heights are simply one-fifth of the scale-lengths and for NGC~3521 and NGC~3621 respectively are 1.9 and 1.5~kpc. These values are several times larger than the fitted scale-heights of 0.25 and 0.5 kpc respectively. If these suggested scale-heights were enforced, it would not be possible to achieve a good fit to the rotation curves. Therefore, it is vital to have a separate measurement of the stellar velocity dispersion to put an orthogonal limit on the scale-heights, as shown by \cite{puglielli10} in the case of NGC~6503 and \cite{bershady11}. \cite{bershady11} used the Disk Mass Survey to demonstrate that the magnitude of vertical velocity dispersions led to the inference that galaxy disks must be submaximal. Maximum disk simply means the fit to the rotation curve with the highest $M/L$ possible and the least amount of dark matter, such that all the rotation velocity is attributable to the disk in the central part. It would be interesting to see if this holds in MOND for which the disk, by definition, must be maximal.

\cite{banerjee11} made an analysis of the scale-heights of the gas dominated THINGS galaxies in Newtonian gravity and it is essential that a similar review is made of the same galaxies in MOND.

\subsection{The external field effect}
\protect\label{sec:efe}
 External gravitational fields (see forthcoming paper), usually the result of nearby galaxies or clusters, can cause a suppression of the boost to gravity due to MOND. This is particularly likely for low surface brightness galaxies. An external gravitational field has a similar effect to decreasing the $M/L$ and increasing the scale-height, but they are not degenerate given the gaseous mass content and the radial dependence of the circular velocity on scale-height. It is an oft-quoted solution to unknown problems in MOND, but it is important to emphasise that it would have no beneficial influence on NGC~2403 or NGC~2903 since it would simply impose a larger $M/L$, making the situation worse.

\begin{table*}
\begin{tabular}{|l|cc||c|cc|cc|c|}
\hline
\small Galaxy& \small  Measured distance & \small ${\rm Fitted~ distance \over \rm Measured~ distance}$ & \small Stellar (Gas) Mass &  \small Predicted $M/L$   & \small ${\rm Fitted~ M/L \over \rm dS~ M/L}$ & \small Suggested $z_*$ & \small Fitted $z_*$ ($z_g$) & Line-type\\
      & Mpc&  & $10^9\msun$ & 3.6$\mu m$ d-S (Kr)  &   &kpc&kpc& \\

\hline
DDO 154-a & 4.30 $\pm$ 1.07 & 1.085  & 0.026 (0.468)& 0.32 (0.23) & 1.0  & 0.2 & 0.2 (1.5)&Solid \\
DDO 154-b&                 & 0.96 &             &               & 1.0&     & 0.2 (0.65)&Dashed\\
DDO 154-c&                 & 0.9 &             &               & 1.0&     & 0.2 (0.05)&Dotted\\
\hline
NGC 3621-a& 6.64 $\pm$ 0.70 &0.91 & 19.3 (9.58)  & 0.59 (0.42) & 0.88  & 1.5 & 0.5 (1.0)&Solid \\
NGC 3621-b&                 & 0.97 &             &               & 0.76&     & 0.5 (1.0)&Dashed\\
NGC 3621-c&                 & 0.91 &             &               & 0.88&     & 0.05 (1.0)&Dotted\\
\hline
NGC 2903-a& 8.90 $\pm$ 2.20 & 0.8 & 16.2 (6.6)   & 0.61 (0.43) & 4.6  & 0.5 & 0.54 (1.0)&Solid\\
NGC 2903-b&                 & 1.1 &             &               & 2.5 &     & 0.3 (1.0)&Dashed\\
NGC 2903-c&                 & 1.15 &             &               & 1.5&     & 0.2 (1.0)&Dotted\\
\hline
NGC 2403-a& 3.47 $\pm$ 0.29 & 1.19 & 5.13 (3.82)  & 0.39 (0.26) & 1.15   & 0.4 & 0.4 (0.1)&Solid\\
NGC 2403-b&                 & 1.09 &             &               & 1.45&     & 0.5 (0.1)&Dashed\\
NGC 2403-c&                 & 1.19 &             &               & 1.15&     & 0.05 (0.1)&Dotted\\
NGC 2403-d&                 & 1.05 &             &               & 1.2&     & 0.6 (0.1)&Dot-Dashed\\
\hline
NGC 3521-a&10.7  $\pm$ 3.20 & 0.63 & 125.5 (13.0) & 0.73 (0.52) & 1.1  & 1.9 & 0.25 (1.0)&Solid\\
NGC 3521-b&                 & 0.82 &             &               & 0.8&     & 0.25 (1.0)&Dashed\\
\hline
\end{tabular}
\caption{ Here we show the various parameters corresponding to our fitted models that are plotted in Figs~\ref{fig:com154} to \ref{fig:com3521}). For each galaxy we give the measured parameters from the literature and our fitted values. The measured distances come from Walter et al. (2008), who took their value from Freedman et al. (2001), except NGC~2403 which comes from the more recent study of Vink\'{o} et al. (2006). The masses for the stellar and gaseous components come from de Blok et al. (2008) and are correct at the mean of the measured distance and with use of the diet-Salpeter initial mass function. The predicted mass-to-light ratios ($M/L$) also come from de Blok et al. (2008), for which ``d-S'' and ``Kr'' correspond to diet-Salpeter and Kroupa initial mass functions respectively. The suggested stellar scale-heights are merely one-fifth of the radial scale-height.  There is no suggested scale-height for the gaseous distribution. For each galaxy, every separate fit is given a different a letter - a, b, c or d - and these fits correspond to different linetypes in their corresponding figures. Our fitted distances and values for the $M/L$ are normalised to the mean measured distance and the diet-Salpeter IMF respectively. The stellar scale-heights are well constrained, but the gas scale-heights less so. For DDO 154 we fixed the stellar parameters since they had little influence.}
\protect\label{tab:par}
\end{table*}

\section{Conclusion}
Here we have introduced an N-body code that solves the modified Poisson equation of QUMOND (see \citealt{milgrom10}). With it, we fitted the rotation curves of five spiral galaxies from the THINGS survey (\citealt{walter08}), using N-body realisations of the stars and the gas in each galaxy fixed by their surface densities. We allowed the distance, mass-to-light ratio and both the scale-heights of the stellar and gaseous disks to be free parameters, with priors set depending on observational constraints.

We discovered that our best fits were excellent matches to the rotation curves of all five galaxies, except for some minor discrepancies at locations of publicised uncertainty in the observations. 
We displayed contour plots of reduced $\chi^2$ for the various free parameters of all galaxies and found that two galaxies, NGC~2403 and NGC~2903, could only make satisfactory fits if their mass-to-light ratios were larger than the predicted values. The required increase in $M/L$ for NGC~2403 is moderate, but for NGC~2903 it is considerable. Interestingly, the dark matter fits to these rotation curves given by \cite{deblok08}, which can be seen in their Tables 5 and 6, also require larger $M/L$. For NGC~2403, a value 50\% larger than the predicted diet-Salpeter value and for NGC~2903 more than double was required, which is exactly what we have found. 

If this were an isolated incident, it could have less relevance, but there is a growing preference for stellar dominated galaxies to require much larger $M/L$s than predicted by stellar population models from typical initial stellar mass functions. Large $M/L$s are also required for certain dwarf spheroidal galaxies surrounding the Milky Way, in particular Carina, Sextans and Draco (see \citealt{angus08,serra10}) and  a large fraction of the early type galaxies studied by \cite{sandnoord}. The MOND fits to the dynamics are in general still excellent, but a solution must be found to explain why certain galaxies can have $M/L$ described by a Kroupa or diet-Salpeter initial mass function and other galaxies need one up to twice as large. Both the MOND and dark matter fits to the rotation curves of NGC~2403 and NGC~2903 suggest there is a problem with the stellar population synthesis models of these galaxies. Furthermore, the distance required for NGC~3521 is considerably lower than the mean, but this is acceptable due to the large uncertainty. A revised distance with tighted error bars could be very revealing. These issues need to be resolved before we can say that MOND provides good fits to all five galaxies with reasonable parameters.

The surprising result was that the MOND fits put a tight constraint on the stellar scale-heights. The best fit scale-heights of all four stellar dominated galaxies were found to be between 0.25~kpc and 0.55~kpc and were strongly constrained to be larger than 0 and less than 1~kpc. For two galaxies, NGC 2403 and 2903, the best fit stellar scale-heights were very close to the typical scale-height used, which is one fifth of the radial scale-length. On the other hand, NGC 3521 and NGC 3621 have best fit stellar scale-heights that are far lower than the suggested scale-height insofar as if the scale-height was fixed to this suggested value, a good fit to the inner data points would not be achievable. Similarly, the only gas dominated galaxy we studied, DDO~154, can achieve a good fit only if its gas scale-height is larger than $\sim$0.6~kpc. In fact the quality of the fit improves with increasing scale-height.

Clearly, the use of both stellar and gas scale-heights as free parameters in MOND fits to galaxy rotation curves is imperative. Line of sight velocity dispersions of both components in a sample of similar face on galaxies should be used as a sanity check on their values since in MOND the rotation curves and vertical velocity dispersions of both components are bound together to the mass distribution.

\section{acknowledgments} GWA, KvdH and WJGdB's research is supported by the University of Cape Town and the National Research Foundation of South Africa. BF acknowledges the support of the AvH foundation. G. G. is a postdoctoral fellow with the National Science Fund (FWO-Vlaanderen). The authors thank the THINGS collaboration for providing their data.


\begin{thebibliography}{37}
\expandafter\ifx\csname natexlab\endcsname\relax\def\natexlab#1{#1}\fi

\bibitem[{{Angus}(2008)}]{angus08}
{Angus} G.~W., 2008, \mnras, 387, 1481

\bibitem[{{Angus} \& {Diaferio}(2011)}]{angus11}
{Angus} G.~W., {Diaferio} A., 2011, preprint(ArXiv:1104.5040)

\bibitem[{{Banerjee} {et~al.}(2011){Banerjee}, {Jog}, {Brinks}, \&
  {Bagetakos}}]{banerjee11}
{Banerjee} A., {Jog} C.~J., {Brinks} E., {Bagetakos} I., 2011,
  preprint(ArXiv:1103.4494)

\bibitem[{{Bekenstein} \& {Milgrom}(1984)}]{bekenstein84}
{Bekenstein} J., {Milgrom} M., 1984, \apj, 286, 7

\bibitem[{{Bershady} {et~al.}(2011){Bershady}, {Martinsson}, {Verheijen},
  {Westfall}, {Andersen}, \& {Swaters}}]{bershady11}
{Bershady} M.~A., {Martinsson} T.~P.~K., {Verheijen} M.~A.~W., {Westfall}
  K.~B., {Andersen} D.~R., {Swaters} R.~A., 2011, \apjl, 739, L47

\bibitem[{{Binney} \& {Tremaine}(2008)}]{bt08}
{Binney} J., {Tremaine} S., 2008, {Galactic Dynamics: Second Edition}. Galactic
  Dynamics: Second Edition, by James Binney and Scott Tremaine.~ISBN
  978-0-691-13026-2 (HB).~Published by Princeton University Press, Princeton,
  NJ USA, 2008.

\bibitem[{{Bottema}(1993)}]{bottema93}
{Bottema} R., 1993, \aap, 275, 16

\bibitem[{{Brada} \& {Milgrom}(1995)}]{brada95}
{Brada} R., {Milgrom} M., 1995, \mnras, 276, 453

\bibitem[{{Brada} \& {Milgrom}(1999)}]{brada99b}
---, 1999, \apj, 519, 590

\bibitem[{{Brada} \& {Milgrom}(2000{\natexlab{a}})}]{brada00b}
---, 2000{\natexlab{a}}, \apj, 541, 556

\bibitem[{{Brada} \& {Milgrom}(2000{\natexlab{b}})}]{brada00a}
---, 2000{\natexlab{b}}, \apjl, 531, L21

\bibitem[{{Carignan} \& {Purton}(1998)}]{carpur98}
{Carignan} C., {Purton} C., 1998, \apj, 506, 125

\bibitem[{{de Blok} {et~al.}(2008){de Blok}, {Walter}, {Brinks},
  {Trachternach}, {Oh}, \& {Kennicutt}}]{deblok08}
{de Blok} W.~J.~G., {Walter} F., {Brinks} E., {Trachternach} C., {Oh} S.,
  {Kennicutt} R.~C., 2008, \aj, 136, 2648

\bibitem[{{Famaey} \& {Binney}(2005)}]{fb05}
{Famaey} B., {Binney} J., 2005, \mnras, 363, 603

\bibitem[{{Famaey} \& {McGaugh}(2011)}]{famaey12}
{Famaey} B., {McGaugh} S., 2011, ArXiv e-prints

\bibitem[{{Gentile} {et~al.}(2011){Gentile}, {Famaey}, \& {de
  Blok}}]{gentile11}
{Gentile} G., {Famaey} B., {de Blok} W.~J.~G., 2011, \aap, 527, A76

\bibitem[{{Kennicutt} {et~al.}(2003){Kennicutt}, {Armus}, {Bendo}, {Calzetti},
  {Dale}, {Draine}, {Engelbracht}, {Gordon}, {Grauer}, {Helou}, {Hollenbach},
  {Jarrett}, {Kewley}, {Leitherer}, {Li}, {Malhotra}, {Regan}, {Rieke},
  {Rieke}, {Roussel}, {Smith}, {Thornley}, \& {Walter}}]{kenn03}
{Kennicutt} Jr. R.~C., {Armus} L., {Bendo} G., {Calzetti} D., {Dale} D.~A.,
  {Draine} B.~T., {Engelbracht} C.~W., {Gordon} K.~D., {Grauer} A.~D., {Helou}
  G., {Hollenbach} D.~J., {Jarrett} T.~H., {Kewley} L.~J., {Leitherer} C., {Li}
  A., {Malhotra} S., {Regan} M.~W., {Rieke} G.~H., {Rieke} M.~J., {Roussel} H.,
  {Smith} J.-D.~T., {Thornley} M.~D., {Walter} F., 2003, PASP, 115, 928

\bibitem[{{Llinares}(2011)}]{llinares11}
{Llinares} C., 2011, PhD thesis, ~Kapteyn Astronomical Institute.~Groningen,
  ISBN 978-90-367-4760-8

\bibitem[{{Llinares} {et~al.}(2008){Llinares}, {Knebe}, \& {Zhao}}]{llinares08}
{Llinares} C., {Knebe} A., {Zhao} H., 2008, \mnras, 391, 1778

\bibitem[{{Londrillo} \& {Nipoti}(2009)}]{londrillo09}
{Londrillo} P., {Nipoti} C., 2009, Memorie della Societa Astronomica Italiana
  Supplementi, 13, 89

\bibitem[{{Milgrom}(1983)}]{milgrom83a}
{Milgrom} M., 1983, \apj, 270, 365

\bibitem[{{Milgrom}(1986)}]{milgrom86}
---, 1986, \apj, 302, 617

\bibitem[{{Milgrom}(2010)}]{milgrom10}
---, 2010, \mnras, 403, 886

\bibitem[{{Nipoti} {et~al.}(2008){Nipoti}, {Ciotti}, {Binney}, \&
  {Londrillo}}]{nipoti08}
{Nipoti} C., {Ciotti} L., {Binney} J., {Londrillo} P., 2008, \mnras, 472

\bibitem[{{Nipoti} {et~al.}(2007{\natexlab{a}}){Nipoti}, {Londrillo}, \&
  {Ciotti}}]{nipoti07a}
{Nipoti} C., {Londrillo} P., {Ciotti} L., 2007{\natexlab{a}}, \apj, 660, 256

\bibitem[{{Nipoti} {et~al.}(2007{\natexlab{b}}){Nipoti}, {Londrillo}, \&
  {Ciotti}}]{nipoti07c}
---, 2007{\natexlab{b}}, \mnras, 381, L104

\bibitem[{{Press} {et~al.}(1992){Press}, {Teukolsky}, {Vetterling}, \&
  {Flannery}}]{numrec}
{Press} W.~H., {Teukolsky} S.~A., {Vetterling} W.~T., {Flannery} B.~P., 1992,
  {Numerical recipes in FORTRAN. The art of scientific computing}, {Press,
  W.~H., Teukolsky, S.~A., Vetterling, W.~T., \& Flannery, B.~P. }, ed.

\bibitem[{{Puglielli} {et~al.}(2010){Puglielli}, {Widrow}, \&
  {Courteau}}]{puglielli10}
{Puglielli} D., {Widrow} L.~M., {Courteau} S., 2010, \apj, 715, 1152

\bibitem[{{Sanders} \& {Noordermeer}(2007)}]{sandnoord}
{Sanders} R.~H., {Noordermeer} E., 2007, \mnras, 379, 702

\bibitem[{{Serra} {et~al.}(2010){Serra}, {Angus}, \& {Diaferio}}]{serra10}
{Serra} A.~L., {Angus} G.~W., {Diaferio} A., 2010, \aap, 524, 16

\bibitem[{{Tiret} \& {Combes}(2007)}]{tcevol}
{Tiret} O., {Combes} F., 2007, \aap, 464, 517

\bibitem[{{Tiret} \& {Combes}(2008{\natexlab{a}})}]{tc08}
---, 2008{\natexlab{a}}, preprint(ArXiv:0803.2631)

\bibitem[{{Tiret} \& {Combes}(2008{\natexlab{b}})}]{tiret08a}
---, 2008{\natexlab{b}}, in Astronomical Society of the Pacific Conference
  Series, Vol. 396, Astronomical Society of the Pacific Conference Series,
  {J.~G.~Funes \& E.~M.~Corsini}, ed., p. 259

\bibitem[{{Walter} \& {Brinks}(1999)}]{walter99}
{Walter} F., {Brinks} E., 1999, \aj, 118, 273

\bibitem[{{Walter} {et~al.}(2008){Walter}, {Brinks}, {de Blok}, {Bigiel},
  {Kennicutt}, {Thornley}, \& {Leroy}}]{walter08}
{Walter} F., {Brinks} E., {de Blok} W.~J.~G., {Bigiel} F., {Kennicutt} R.~C.,
  {Thornley} M.~D., {Leroy} A., 2008, \aj, 136, 2563

\bibitem[{{Wu} {et~al.}(2009){Wu}, {Zhao}, {Wang}, {Llinares}, \&
  {Knebe}}]{wu09}
{Wu} X., {Zhao} H., {Wang} Y., {Llinares} C., {Knebe} A., 2009, \mnras, 396,
  109

\bibitem[{{Zhao} \& {Famaey}(2010)}]{zhao10}
{Zhao} H., {Famaey} B., 2010, \prd, 81, 087304

\end{thebibliography}
\end{document}